\documentclass[12pt,lettersize]{article}

\usepackage{amssymb}
\usepackage{amsmath,amsthm}
\usepackage{mathrsfs}
\usepackage{bbm}
\usepackage{hyperref}
\usepackage{array,longtable}
\usepackage[dvips]{graphicx}
\usepackage[dvips]{color}

\newcommand{\Ds}{\mathscr{D}}

\newcommand{\Rh}{\mathbb R}

\newcommand{\Zh}{\mathbb Z}

\newcommand{\Ac}{\mathcal{A}}

\newcommand{\Nc}{\mathcal{N}}
\newcommand{\Jc}{\mathcal{J}}

\newcommand{\Hc}{\mathcal{H}}

\newcommand{\Oc}{\mathcal{O}}

\newcommand{\pd}{\partial}

\newcommand{\oz}{\overline{z}}

\newcommand{\Tr}{\mathop{\mathrm{Tr}}\nolimits}

\newcommand{\Oplus}{\mathop{\oplus}\limits}
\newcommand{\CS}{\mathop{\textsl{CS}\,}\nolimits}

\renewcommand{\Im}{\mathop{\mathrm{Im}}\nolimits}
\renewcommand{\Re}{\mathop{\mathrm{Re}}\nolimits}

\renewcommand{\theequation}{\arabic{section}.\arabic{equation}}

\begin{document}

\title{
\begin{flushright}
{\small hep-th/0505235}
\end{flushright}
\vspace{1cm}
\textbf{Classification of abelian  spin Chern-Simons theories  }}
\author{
{Dmitriy M.~Belov
\; and
\;Gregory W.~Moore}
\vspace{5mm}
\\
\emph{
Department of Physics, Rutgers University}
\\
\emph{136 Frelinghuysen Rd., Piscataway, NJ 08854, USA}
}

\date {~}

\maketitle
\thispagestyle{empty}

\begin{abstract}
We derive a simple classification of quantum   spin
Chern-Simons
theories with gauge group $T=U(1)^N$.
While the  classical Chern-Simons theories
are classified by an integral lattice the quantum
theories are classified differently. Two quantum
theories are equivalent if they have the same invariants
on 3-manifolds with spin structure, or equivalently if they lead to equivalent
projective representations of the modular group.
We prove the quantum theory  is completely
determined by three invariants which can be constructed
from the data in the classical action.
We comment on implications for the  classification of
fractional quantum Hall fluids.
\end{abstract}

\vspace{1cm}
$~~$ May  2005

\clearpage

\tableofcontents

\section{Introduction}
\label{sec:intro}
\setcounter{equation}{0}

This paper is about three-dimensional Chern-Simons gauge theories.
These theories have a variety of applications in topology,
conformal field theory, the AdS/CFT correspondence, and in the fractional quantum
Hall effect (FQHE). Here we will investigate quantum equivalences between
Chern-Simons theories with gauge group $U(1)^N$. In
\cite{Witten:2003ya} Witten pointed out that nontrivial quantum
equivalences between such theories exist. This raises the question
of the classification of such theories. The present paper answers
that question.

\paragraph{Classical (spin) Chern-Simons action.}
The classification of abelian Chern-Simons theories was mentioned in
\cite{Gukov:2004id} but, especially in view of the potential
applications to the FQHE, the problem of real interest is the classification
of {\it spin} Chern Simons theories, i.e. Chern-Simons theories at
half-integer level. We now briefly recall these theories, which were
originally introduced in \cite{DW}.

For any compact gauge group $G$ the Chern-Simons actions are classified by
an element $k\in H^4(BG,\Zh)$ \cite{DW,Freed:1992vw}. For $G=U(1)$,
$H^4(BG,\Zh) \cong \Zh$, and the element is simply the  integer $k$
appearing in the familiar expression for the $U(1)$ Chern-Simons
action on a 3-manifold $X_3$:
\begin{equation*}
e^{2\pi i\, CS(A)}=\exp\Bigl[ 2\pi i k   \int_{X_3} A \wedge dA\Bigr]
\end{equation*}
(In this paper the connection is normalized so that $F=dA$ has
integral periods.)
As usual the expression is defined by choosing an
extension to a bounding 4-manifold $Z_4$ and defining
\begin{equation}
e^{2\pi i CS(A)}:=\exp\Bigl[ 2\pi i k\int_{Z_4} \tilde F\wedge \tilde F\Bigr].
\end{equation}
where $\tilde F$ is the fieldstrength of an extending connection.
This expression does not depend on the extension
provided that $k$ is integral. This follows since
 $\int F^2$ on a closed 4-manifold is an integer.
 Since it can be any integer the minimal value of $k$ is $1$. Now, if we
endow $X_3$ with a spin structure and require that the bounding
4-manifold $Z_4$ have a compatible spin structure then
$\int F \wedge F $ is always an {\it even} integer and hence
$k$ can be half-integral. This defines \textit{a spin} (or half-integral) $U(1)$ Chern-Simons
theory.

More generally, if the gauge group is a torus $T= U(1)^N$
then it is convenient to think of $H^4(BT,\Zh)$
as the  space of integral \textit{even} symmetric bilinear forms.
Indeed, choosing
a basis $c_1^\alpha$ for $H^2(BT,\Zh)$ an element in
$H^4(BT,\Zh)$ can be written as
$M_{\alpha\beta}\, c_1^\alpha \cup c_1^\beta$ corresponding to the
Chern-Simons action
\begin{equation}
e^{2\pi i\, CS(A)}=\exp\Bigl[ i \pi     \int_{Z_4} M_{\alpha\beta}
\tilde{F}^\alpha \wedge \tilde{F}^\beta\Bigr]
=\exp\Bigl[ i \pi     \int_{Z_4} M_{\alpha\alpha}\tilde{F}^\alpha \wedge \tilde{F}^\alpha
+\sum_{\alpha<\beta}2M_{\alpha\beta}\tilde{F}^\alpha \wedge \tilde{F}^\beta\Bigr].
\label{CSaction}
\end{equation}
This expression does not depend on the extension provided that
$M_{\alpha\alpha}$ is an even integer and $M_{\alpha\beta}$ for $\alpha\ne\beta$
is an arbitrary integer. Thus the matrix $M_{\alpha\beta}$ indeed
defines an integral even symmetric bilinear form.
Every integral symmetric bilinear form defines a lattice.
Thus the Chern-Simons actions with the gauge group $U(1)^N$ are
classified by \textit{even} integral lattices of rank $N$.

Now, if we endow $X_3$ with a spin structure and require that the bounding
4-manifold $Z_4$ have a compatible spin structure then
$\int F^{\alpha} \wedge F^{\alpha} $ is always an {\it even} integer and hence
$M_{\alpha\alpha}$ in \eqref{CSaction} can be any integer, not
necessary even. Thus \textit{spin} Chern-Simons theories
are classified by   integral symmetric bilinear forms, or
more invariantly by integral (generically odd) lattices. Having an integral
symmetric bilinear form $K_{\alpha\beta}$ we define an action
for \textit{spin} Chern-Simons theory by
\begin{equation}
e^{2\pi i\, CS^{\text{spin}}_{\pd Z_4}(A)}=\exp\Bigl[ i \pi     \int_{Z_4} K_{\alpha\beta}
\tilde{F}^\alpha \wedge \tilde{F}^\beta\Bigr].
\label{CSaction}
\end{equation}
We denote by $\Lambda$ the lattice $\Lambda=\Zh\langle e_1,\dots,e_N\rangle$
with the bilinear form $(e_{\alpha},e_{\beta})=K_{\alpha\beta}$.
In general the spin Chern-Simons theories are classified by a group in a certain cohomology theory denoted
by $E^4(BG)$ \cite{Freed:1992vw}. In the case at hand
$E^4(BT)$
is isomorphic to the space of arbitrary integral
symmetric bilinear forms.
Further details about spin Chern-Simons theories can be found
in section 2. We will assume the bilinear form is
nondegenerate and also that the induced form on $\Lambda^*/\Lambda$ is
nondegenerate.

\paragraph{The quantum equivalence.}
While the classification of   abelian spin Chern-Simons actions is
the classification of integral lattices, there are, as we have
mentioned, quantum equivalences between theories with different
classical actions. We will define two theories to be equivalent if they
have the same 3-manifold invariants. Equivalently, we declare them
to be equivalent if the ``conformal blocks'' (which can be nonholomorphic)
transform in equivalent projective representations of the modular group.
The main result of this paper is a
 theorem on the classification of the quantum theories under this
equivalence relation.
Our result says that up to isomorphism  the quantum theory is
determined  by the following $3$ invariants of $\Lambda$ subject to
a constraint:

\noindent\textbf{Data:}
\begin{enumerate}
\item $\sigma \mod 24 $ the signature of the
bilinear form on $\Lambda$ modulo $24$.

\item Discriminant group $\mathscr{D}=\Lambda^*/\Lambda$ with
the bilinear form $b:\mathscr{D}\times \mathscr{D}\to \mathbb{Q}/\Zh$
inherited from the bilinear form on $\Lambda^*$.

\item The equivalence class of a quadratic refinement $[q(\gamma)]$
of the bilinear form $b$
on $\mathscr{D}$. The quadratic refinement is a map $q:\mathscr{D}\to \mathbb{Q}/\Zh$
such that for any elements $\gamma_1,\gamma_2$ of $\mathscr{D}$ we have
\begin{equation}
q(\gamma_1+\gamma_2)-q(\gamma_1)-q(\gamma_2)+q(0)=b(\gamma_1,\gamma_2).
\label{q.ref}
\end{equation}
We say that two quadratic refinements $q_1$ and $q_2$
are equivalent if there exists
$\Delta\in \mathscr{D}$  such that for all $\gamma\in \mathscr{D}$
\begin{equation}
q_1(\gamma)=q_2(\gamma-\Delta).
\label{equiv}
\end{equation}
\end{enumerate}
\noindent\textbf{Constraint:}
The constraint on the data follows from the Gauss-Milgram sum formula \cite{Milgram,Milnor}.
Note that if $q$ is a quadratic refinement then so is
$q+c$ for any constant $c$. We fix $[q]$ by the requirement that:
\begin{equation}
|\mathscr{D}|^{-1/2}\sum_{\gamma\in \mathscr{D}}
e^{2\pi i q(\gamma)}=e^{2\pi i \sigma/8}.
\label{constraint}
\end{equation}
Notice that the constraint does not depend on a particular choice
of representative $q$ of the equivalence class $[q]$.

For an arbitrary lattice $\Lambda$ the constraint \eqref{constraint}
is very restrictive and defines a unique equivalence class of
  quadratic refinements. The construction goes as follows.
Denote by $W_2$ a representative
of the characteristic class $[W_2]\in\Lambda^*/2\Lambda^*$:
for all $\lambda\in\Lambda$ it satisfies $(\lambda,\lambda)=W_2(\lambda)\mod 2$.
Then the quadratic refinement satisfying \eqref{constraint} may be
represented by
\begin{equation}
q_{W_2}(\gamma)=\frac12(\gamma,\gamma-W_2)+\frac18(W_2,W_2)
\mod 1
\end{equation}
where $(\cdot,\cdot)$ denotes the bilinear form on $\Lambda^*$.
The equivalence class of the quadratic refinement
is filled out by choosing
different representatives $W_2$ of the characteristic class $[W_2]$. Evidently this set of
quadratic refinements satisfies the equivalence relation \eqref{equiv}.
%

This result raises the natural question of whether there is a converse result.
Using classical theorems about lattices (see \cite{Conway,Nikulin})
one can prove that every quartet $(\sigma\mod 24,\,\mathscr{D},\,b,\,[q])$
satisfying equation \eqref{constraint} is realized by some lattice.
Moreover if there exists a representative $q$ of  $[q]$ such that $q(0)=0$ then
the quartet is realized by an \textit{even} lattice.
Thus we have the following

\vspace{3mm}
\noindent\textbf{Classification of quantum Chern-Simons theories:}
\begin{enumerate}
\item[I.]
The set of quantum
\textit{spin} Chern-Simons theories is in one-to-one
correspondence with the quartets $(\sigma\mod 24,\,\mathscr{D},\,b,\,[q])$
subject to the Gauss-Milgram constraint \eqref{constraint}.

\item[II.]
The set of quantum \textit{integral} Chern-Simons
theories is in one-to-one correspondence with
the quartets subject to the Gauss-Milgram constraint such that
there exists a representative $q$ of the equivalence class $[q]$ with $q(0)=0$.
\end{enumerate}
This is the main result of this paper.

We prove the above result by careful construction of the groundstate
wavefunction (or conformal blocks) of the theory on a compact Riemann
surface of genus $g$. This gives a very explicit projective representation of
the modular group. The level $k$ $U(1)$ Chern-Simons theory was carefully
constructed in \cite{Manoliu}. In the language of this paper,
ref. \cite{Manoliu} investigates the theory corresponding to the
quadratic form $\left(\begin{smallmatrix}
k & 0 \\
0 & -1
\end{smallmatrix}\right)$.

The space of quantum Chern-Simons theories has
a structure of monoid (a semigroup with an identity element).
The sum operation is defined by
\begin{equation*}
(\sigma_1\mod 24,\,\mathscr{D}_1,\,b_1,\,[q_1])
\oplus (\sigma_2\mod 24,\,\mathscr{D}_2,\,b_2,\,[q_2])
=(\sigma_1+\sigma_2\mod 24,\,\mathscr{D}_1\oplus\mathscr{D}_2,\,b_{12},\,[q_{12}])
\end{equation*}
where the bilinear form $b_{12}$ and the quadratic refinement $q_{12}$
are defined by
\begin{align*}
b_{12}(x_1\oplus x_2,\,y_1\oplus y_2)&=b_1(x_1,x_2)+b_2(x_2,y_2)\mod 1
\\
q_{12}(x_1\oplus x_2)&=q_1(x_1)+q_2(x_2)\mod 1.
\end{align*}
The quadratic refinement defined in this way
satisfies the Gauss-Milgram constraint with the signature $(\sigma_1+\sigma_2)\mod 8$.
The identity element is the trivial quartet.
One can ask whether the space of quantum Chern-Simons
theories has a group structure. The answer appears to be \textit{no}.
See   the end of section~\ref{sec:sum} for more details.

\paragraph{Fractional quantum Hall effect.}
We hope that our result will be useful in the theory of the
fractional quantum Hall effect (FQHE). It has been known for some time
\cite{Read,Frohlich:1991wb,Wen,Zee,Frohlich:1994nk} that the classification of quantum
Hall fluids is related to the classification of quantum Chern-Simons
theories. Most of these discussions have been phrased in terms
of the classification of lattices. In section~\ref{sec:QFHE} we recall some
aspects of the relation to the FQHE. One consequence of our result is
that an effective theory with invariants 1,2,3 specified above  can give
the following fractional Hall conductivities $\sigma_H$ in the ground state:
\begin{equation}
\sigma_H\!\mod 8 \in\Sigma_H\quad \text{where}\quad
\Sigma_H=\{8\,q(\gamma)\!\mod 8\,|\,\gamma\in \mathscr{D}\}.
\end{equation}
The electric charges (modulo $1$) of the vortices $\gamma\in \mathscr{D}$
are given by $q(-\gamma)-q(\gamma)$.
An example of this construction is given in section~7.

\vspace{5mm}
The paper is organized as follows. In section~\ref{sec:intro}
we review the construction of spin Chern-Simons actions.
In section~\ref{sec:ham} we
consider Hamiltonian quantization of the spin Chern-Simons  theory.
In section~\ref{sec:gauss} we formulate the quantum Gauss law.
In section~\ref{sec:wave} we derive the physical wave function and
discuss its properties. Section~\ref{sec:sum} summarizes the main theorem
and section~\ref{sec:QFHE} discusses the fractional quantum Hall effect.

\section{Spin Chern-Simons theory}
\label{sec:spinCS}
\setcounter{equation}{0}

First let us recall the construction of Chern-Simons
theories in $3$ dimensions. Let $G$ be a compact Lie group
and let $P\to X_3$ be a principal $G$-bundle with connection $A$.
Choose a real representation $\rho$ and let $E_{\rho}\to X_3$
be the associated bundle.
We will denote a representation of the Lie algebra $\mathfrak{g}$ by
the same symbol $\rho$.
If $G$ is a connected simply connected compact Lie group then
the bundle $E_{\rho}$ is topologically trivial. Thus $A$ is a globally well defined
Lie algebra valued $1$-form on $X_3$.
In this case the Chern-Simons functional can be defined by the formula
\begin{equation}
e^{2\pi i\CS_{X_3}(A)}=e^{2\pi i\int_{X_3}-\frac{1}{8\pi^2}\Tr_{\rho} (A\wedge dA+\frac23 A^3)}.
\label{CSnaive}
\end{equation}
$\Tr_{\rho}$ is a $G$-invariant quadratic form on
the Lie algebra $\mathfrak{g}$. We temporarily refer to standard
normalization of gauge field.

If $E_{\rho}$ is a nontrivial $G$ bundle then \eqref{CSnaive} cannot
be used to define CS theory. However if there exists a $4$-manifold $Z_4$
which bounds $X_3$ and an extension of the bundle $E_{\rho}$ over $X_3$
to the bundle $\tilde{E}_{\rho}$ over $Z_4$ then we can define the Chern-Simons functional by
\begin{equation}
e^{2\pi i\CS_{\pd Z_4}(A)}=e^{2\pi i\int_{Z_4}(-\frac{1}{8\pi^2}\Tr_{\rho} \tilde{F}\wedge \tilde{F})}.
\label{CSZnonspin}
\end{equation}
Here $\tilde{F}$ is the curvature of the extended connection $\tilde{A}$.
Notice that if the bundle extends then the connection can always
be extended via a partition of unity.
The term $e^{2\pi i\CS_{X_3}(A)}$ appears in the path integral and thus
it must be single-valued. This requires the integral
of $-\frac{1}{8\pi^2}\Tr_{\rho} \tilde{F}\wedge \tilde{F}$ over an
arbitrary closed $4$-manifold $Z_4$
to be integral. In virtue of its relations to the first Pontragin
class it is indeed true.

Recall that there is a universal bundle $G\to EG\to BG$.
To construct a $G$ bundle over $X_3$ one has to find
a classifying map form $X_3$ to $BG$, and
then take a pullback of the universal bundle. Thus the question about
existence of $Z_4$ which bounds $X_3$ such that the bundle $E$ extends is
the question of whether the oriented cobordism group $\Omega_3(BG)$ is zero or not.

Recall that any compact oriented $3$-manifold admits a spin structure \cite{Stipsicz}.
If we choose a spin structure on $X_3$ we can modify definition \eqref{CSZnonspin}
by requiring that the spin structure on $X_3$ must be extendable to $Z_4$.
Hence the $4$-manifold $Z_4$ must also be a spin manifold.
In this case the Pontragin class $p_1(E_{\rho})$ is actually
divisible by $2$ (This follows from the index
theorem and the quaternionic structure of the spin bundle).
Thus we can define a square root
from the Chern-Simons functional
\begin{equation}
e^{2\pi i CS^{\text{spin}}_{\pd Z_4}(A)}=
e^{i\pi \int_{Z_4}(-\frac{1}{8\pi^2}\Tr_{\rho} \tilde{F}\wedge \tilde{F})}.
\label{CSZ}
\end{equation}
The theory defined by this action is called a \textit{spin} Chern-Simons theory.
The existence of a bounding spin $4$-manifold $Z_4$ is
related to the spin cobordism group
 $\Omega_3^{spin}(BG)$.

If $a$ is a globally well defined Lie algebra valued one form then
we have the following variational formula for the spin Chern-Simons functional:
\begin{equation}
\CS^{\text{spin}}_{X_3}(A+a)-\CS^{\text{spin}}_{X_3}(A)=-\frac{1}{8\pi^2}\int_{X_3}\Tr_{\rho}(F\wedge a)
-\frac{1}{16\pi^2}\int_{X_3}\Tr_{\rho}\Bigl(a\wedge da+\frac23a^3\Bigr)
\mod 1.
\label{varF}
\end{equation}

\subsection{Abelian spin Chern-Simons theory}
\label{sec:aCS}
For the abelian  group $T=U(1)^N$ it happens that
both cobordism groups
$\Omega_3(BT)$ and $\Omega_3^{spin}(BT)$ vanish\footnote{
We would like to thank Peter Landweber and Robert Stong for explaining
this result to us.}.

The spin Chern-Simons theory is constructed as follows.
Consider an oriented compact 3-dimensional manifold $X_3$ without boundary.
Let $P^{\alpha}\to X_3$
be  oriented principal $SO(2)$ bundles corresponding to the $\alpha$-th
$U(1)$ group in $T$.
The Lie algebra of $SO(2)$
contains one element $\mathfrak{t}=\left(\begin{smallmatrix} 0 & 1\\ -1& 0
\end{smallmatrix}\right)$.
We denote by $A^{\alpha}\otimes \mathfrak{t}$ a connection on $P^{\alpha}$.
In the abelian case we find it convenient to rescale the connection
$A_{\alpha}$ by $2\pi$ so that its curvature $F^{\alpha}$ has
integer periods. We henceforth use this convention.
To construct an arbitrary real $T$ bundle $E_{\rho}$ one has to choose
a real representation $\rho$ of the group $T$.

Any real irreducible representation of $T$ is
two dimensional. We denote $R(t)$ the following matrix
\begin{equation}
R(t)=\begin{pmatrix}
\cos 2\pi t & \sin 2\pi t
\\
-\sin 2\pi t & \cos 2\pi t
\end{pmatrix}.
\end{equation}
Then an arbitrary real irreducible representation
is of the form $R(v_{\alpha}x^{\alpha})$ where
$v_{\alpha}\in \Zh$ and $x^{\alpha}\in\Rh/\Zh$ parameterizes the angle
in the $\alpha$-th $U(1)$ group.
Any virtual real representation $\rho$ of $T$ can be
written as
\begin{equation*}
\rho(x)=\Oplus_{v} m_v\,R(v_{\alpha}x^{\alpha})
\end{equation*}
where $m_v\in \Zh$  is the multiplicity of the irreducible representation $R(v_{\alpha}x^{\alpha})$
and we assume $m_v=0$ for all but finitely many $v$.
Usually multiplicities $m_v$ are positive integers, however here
we consider virtual representations and  thus multiplicities can
also be negative.
It is easy to see that the curvature is
\begin{equation}
\rho(F)=\Oplus_{v} m_v\,(v_{\alpha}F^{\alpha})\otimes\mathfrak{t}
\quad\Rightarrow\quad
-\frac{1}{2}\Tr_{\rho} (F\wedge F)=\sum_v m_v v_{\alpha}v_{\beta}\,F^{\alpha}\wedge F^{\beta}.
\label{cw}
\end{equation}
Using \eqref{CSZ} one finds that the spin Chern-Simons functional is
\begin{equation}
e^{2\pi i \CS^{\text{spin}}_{X_3}(A)}=
e^{i\pi \int_{Z_4}K_{\alpha\beta}\tilde{F}^{\alpha}\wedge \tilde{F}^{\beta}}
\quad\text{where}\quad
K_{\alpha\beta}=\sum_{v}m_vv_{\alpha}v_{\beta}.
\label{oK}
\end{equation}
This expression coincides with \eqref{CSaction} which was obtained
in introduction using simpler arguments.

It is interesting to notice that an arbitrary representation of $U(1)^N$
defines a spin Chern-Simons action but not an integral Chern-Simons
action.  An integral Chern-Simons action is defined only by very special representations
$\rho_e$:
they are generated by $R(v_{\alpha}x^{\alpha})\oplus v_{\beta}\, R(x^{\beta})$.
For these types of representations we have
\begin{equation*}
-\frac{1}{2}\Tr_{\rho_e} [F\wedge F]
=\sum_v m_v v_{\alpha}(v_{\alpha}+1)\,[F^{\alpha}\wedge F^{\alpha}]
+2\sum_{\alpha<\beta}\sum_v m_v v_{\alpha}v_{\beta}\,[F^{\alpha}\wedge F^{\beta}]
\in H^4(BT,\Zh).
\end{equation*}
The corresponding matrix $M_{\alpha\beta}$ has even integers
on the diagonal, and thus it defines an integral Chern-Simons action (see introduction).

\subsubsection{Dependence on a choice of spin structure}
The action \eqref{oK} depends on the choice of spin structure.
The conceptual explanation for this is given in the next subsection.
In the thesis of \cite{Jenquin} it is shown that
for general compact group $G$ spin Chern-Simons actions
are associated with elements of $E^4(BG)$, which fits in
\begin{equation}
0\to H^4(BG,\Zh)\to E^4(BG)\stackrel{w_2}{\longrightarrow} H^2(BG,\Zh_2).
\end{equation}
For the group $T=U(1)^N$ we
choose a basis $\{c_1^{\alpha}\}$ in $H^2(BT,\Zh)$ and write
an element  $e^4(BT)=K_{\alpha\beta}\,c_1^{\alpha}\cup c_1^{\beta}$
where $K$ is a matrix defined in \eqref{oK}.
The projection $w_2$ onto $H^2(BT,\Zh)$ is just the mod $2$ reduction
of the element \eqref{cw} of $H^2(BT,\Zh)$: $w_2(e^4)=\sum_v m_v v_{\alpha} w_2^{\alpha}$
where $w_2^{\alpha}$ are the Stiefel-Whitney
 classes of $P^{\alpha}$.

\begin{figure}[t]
\centering
\includegraphics[width=380pt]{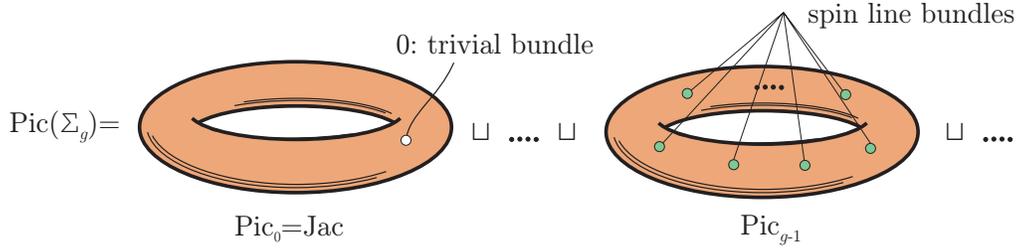}
\caption{The Picard group is a disjoint union of the torii.
The spin line bundles are represented by $2^{2g}$ points in the
$g-1$ component of the Picard group. There is no natural origin
of $\mathrm{Pic}_{g-1}$.}
\label{fig:picard}
\end{figure}
The space $\mathrm{spin}(X)$ of the spin structures of $X$ is
a principal homogenous space for the group of
translations  $H^1(X,\frac12\Zh/\Zh)$ (see Figure~\ref{fig:picard}).
The group $H^1(X,\frac12\Zh/\Zh)$ is in one-to-one correspondence with flat real
line bundles on $X$.
If  $S_{\sigma}$ is the spinor bundle corresponding to the spin structure $\sigma$
and $\epsilon\in H^1(X,\frac12\Zh/\Zh)$ then the bundle $S_{\sigma+\epsilon}$
can be identified with $S_{\sigma}\otimes L_{\epsilon}$
where $L_{\epsilon}$ is a line bundle with the flat connection $\epsilon$.
To get the spin structure dependence of \eqref{oK} we use the close relation
between Chern-Simons and the $\eta$-invariant of a Dirac operator
(see next subsection).
The Dirac operator associated to $S_{\sigma+\epsilon}$
is the Dirac operator $D_{\sigma}$ twisted by $\epsilon$.
Thus to change the spin structure
we have to shift $\rho(A)\mapsto\rho(A)+\epsilon\, \mathbbmss{1}_{\rho}\otimes\mathfrak{t}$.
The variational formula \eqref{varF} yields
\begin{equation}
e^{2\pi i \CS^{\text{spin}}_{\sigma+\epsilon}(A)}=e^{2\pi i \CS^{\text{spin}}_{\sigma}(A)}
e^{2\pi i \int_{X_3}(W_2)_{\alpha}F^{\alpha}\wedge \epsilon}
\label{spinChange}
\end{equation}
where $(W_2)_{\alpha}$ is a solution of the equation
\begin{equation*}
(W_2)_{\alpha}=\sum_v m_v v_{\alpha}\mod 2.
\end{equation*}
Equation \eqref{spinChange} does not depend on a particular
choice of solution.
{}From \eqref{oK} it is easy to see that the equation above can be rewritten
in the form
\begin{equation}
(W_2)_{\alpha}=K_{\alpha\alpha}\mod 2.
\label{W2class}
\end{equation}

\subsubsection{Basis independent notation}
Clearly the matrix $K_{\alpha\beta}$ \eqref{oK} depends on a choice of basis
in $H^2(BT,\Zh)$. However two different matrices define the
same classical action if they are related by a conjugation by an element of
$SL(N,\Zh)$. It is therefore convenient to describe the matrix $K_{\alpha\beta}$
in a more invariant language. Recall that any integral matrix defines
an integral lattice which is a free $\Zh$ module
$\Lambda=\Zh\langle e_1,\dots, e_N\rangle$ with the integral valued bilinear form
given by the matrix $(e_{\alpha},e_{\beta})=K_{\alpha\beta}$.
We denote the dual basis in $\Lambda^*$ by $\{\check{e}^{\alpha}\}$.
In this notation the action \eqref{oK} takes the form
\begin{equation}
e^{2\pi i \CS^{\text{spin}}_{X_3}(A)}=\exp\Bigl[i\pi \int_{Z_4}(\tilde{F},\tilde{F})
\Bigr]
\label{CSphase}
\end{equation}
where $\tilde{F}$ the curvature of the extended connection
$\tilde{A}$. In this notation the curvature $F$ takes values in $\Omega^2_{\Lambda}(X_3)$,
the space of two forms with periods in $\Lambda$.

The vector $(W_2)_{\alpha}\check{e}^{\alpha}$ appearing in \eqref{spinChange}
is a representative of the characteristic class $[W_2]\in\Lambda^*/2\Lambda^*$ defined by
the condition that $W_2(\lambda)=(\lambda,\lambda)\mod 2$
for all $\lambda\in\Lambda$. This condition is nothing else but
the basis independent form of equation \eqref{W2class}.
It is easy to see that the characteristic class $[W_2]$ is uniquely determined by
the bilinear form on the lattice.

If $a^{\alpha}$ is a globally well defined $1$-form then we have the following
variational formula
\begin{subequations}
\begin{equation}
e^{2\pi i \CS_{X_3}^{\text{spin}}(A+a)}=e^{2\pi i \CS^{\text{spin}}_{X_3}(A)}\,\exp\Bigl[
2\pi i \int_{X_3}
(F,a)
+
i\pi \int_{X_3} (a,da)
\Bigr].
\label{variat}
\end{equation}

If ${CS}_{\sigma}$ is the
spin
Chern-Simons functional corresponding to the spin structure $\sigma$
and $\epsilon\in H^1(X_3,\frac12\Zh/\Zh)$ then the Chern-Simons
functional corresponding to the spin structure $\sigma+\epsilon$ is\footnote{We
use the notation $\CS_X$ when we wish to emphasize the boundary
manifold $X$ and $\CS_{\sigma}$ when $X$ is understood and we want
to emphasize the dependence on spin structure $\sigma$.}
\begin{equation}
e^{2\pi i \CS^{\text{spin}}_{\sigma+\epsilon}(A)}=e^{2\pi i\int_{X_3}W_2(F)\wedge \epsilon}\,
e^{2\pi i \CS^{\text{spin}}_{\sigma}(A)}.
\label{VF:spin}
\end{equation}
\label{VF}
\end{subequations}
Here $W_2$ is \textit{any} representative of the characteristic class $[W_2]$.

\subsection{Relation to the $\xi$-invariant}
In \cite{Jenquin} Jenquin applied the results of Dai and Freed \cite{Dai:1994kq} on the
$\xi$-invariant of the
Dirac operator to study  spin Chern-Simons theory.  The spin
Chern-Simons action and
the $\xi$ invariant are sections of the same line bundle over the space of connections,
and hence they can be identified up to multiplication by a metric dependent function $f(g)$:
\begin{equation}
e^{2\pi i \CS^{\text{spin}}(A)}=e^{i\pi\xi(D_A)}f(g).
\label{DFCS}
\end{equation}

The Dirac operator $D_A$ on a compact $3$-dimensional manifold $X_3$
is self-adjoint and elliptic. Its spectrum is real and discrete
so one can define
\begin{equation}
\eta_{D_A}(s)=\sum_{\lambda\ne 0}\frac{\mathrm{sign}\lambda}{|\lambda|^s},\qquad
\Re s>3/2
\end{equation}
where the sum ranges over the nonzero spectrum of $D_A$.
This series converges for $s>3/2$ and defines meromorphic
function which is regular at $s=0$. The value $\eta_{D_A}(0)$
is called the $\eta$-invariant of the Dirac operator $D_A$.
The $\xi$ invariant is
\begin{equation}
\xi(D_A)=\frac{1}{2}\bigl(\eta_{D_A}(0)+\dim\ker D_A\bigr).
\end{equation}
On a general odd dimensional manifold $\xi(D_A)\mod 1$ is a smooth
function of geometric parameters such as
the connection and metric. However in $3$ dimensions (and more generally in
$8k+3$ dimensions) it happens that $\xi(D_A)\mod 2$ is also a smooth function of the
geometric parameters.

The variation of the $\xi$-invariant under the change of smooth parameters
follows from the APS index theorem for families. Consider a path $A(t)$
and $g(t)$, $t\in[0,1]$,  in the space of connections and metrics respectively then
\begin{multline}
\xi_{D_A(1)}-\xi_{D_A(0)}=\int_{X_3\times[0,1]}
\left[\hat{A}\Bigl(\frac{\mathcal{R}(t)}{2\pi}\Bigr)\wedge \mathrm{ch}(F_{A(t)})\right]_{(4)}
\\
=-\frac{1}{8\pi^2}\int_{X_3\times [0,1]}
\Tr_{\rho}(F_{A(t)}\wedge F_{A(t)})+\mathrm{rank}(E)\int_{X_3\times [0,1]}\hat{A}
\Bigl(\frac{\mathcal{R}(t)}{2\pi}\Bigr)
\end{multline}
where $\hat{A}$ is the roof-genus, $\mathcal{R}(t)$ is the Riemann tensor
corresponding to the metric $g(t)$. Comparing
this equation with \eqref{CSZ} one sees that
the $\xi$ invariant and the spin Chern-Simons action
are sections of the same line bundle over the space of connections.

\section{Hamiltonian quantization on $X_3=\Sigma_g\times \Rh$}
\label{sec:ham}
\setcounter{equation}{0}
\subsection{Lagrangian}
Consider Maxwell-Chern-Simons theory with gauge group $U(1)^N$ in three dimensions.
Let $P^{\alpha}$ be principal $U(1)$ bundles corresponding to the
$\alpha$-th gauge group in $U(1)^N$.
We denote by $A^{\alpha}$ ($\alpha=1,\dots,N$) the gauge potential ($\Rh$-valued)
on $P^{\alpha}$.  The action function $S$ of our theory is
\begin{equation}
e^{iS}=\exp\Bigl[
-\frac{i}{2e^2}\int_{X_3}\lambda^{-1}_{\alpha\beta}\,F^{\alpha}*F^{\beta}
\Bigr]e^{2\pi i \CS_{X_3}^{\text{spin}}(A^{\alpha})}
\mathrm{Hol}_{A^{\alpha}}(\{C_r,n^{r}_{\alpha}\}).
\label{action}
\end{equation}
$\alpha,\beta=1,\dots,N$,  $\lambda^{\alpha\beta}$
is a symmetric positive definite
matrix with $\det\lambda=1$. It defines
the relative couplings of $U(1)$ gauge potentials.
$r=1,\dots,N_V$ where $N_V$ is the number of vortices,
$\{C_r,n_{\alpha}^{r}\}$ is a set of contours $C_r$ (vortices) together
with representation $n^r_{\alpha}$ of $U(1)^N$.
$\mathrm{Hol}_{A^{\alpha}}(\{C_r,n^{r}_{\alpha}\})$
is the holonomy of the gauge field $A$ around
the contours $C_r$ in the representation $n_r$.
When $A^{\alpha}$ is globally well defined
we can write the last term in \eqref{action} as $e^{2\pi i\,n^r_{\alpha}\oint_{C_r}
A^{\alpha}}$.
$\CS^{\text{spin}}_{X_3}(A^{\alpha})$
is the spin Chern-Simons term which is defined by \eqref{CSphase}.
It is characterized by an integral (generically odd) lattice $\Lambda$.
We choose a basis $\{e_{\alpha}\}$ in $\Lambda$ and denote the bilinear
form $(e_{\alpha},e_{\beta})$ in this basis by $K_{\alpha\beta}$.
The dual basis in $\Lambda^*$ we denote by $\{\check{e}^{\alpha}\}$.
We can think of the integers $n^r_{\alpha}$ appeared in \eqref{action}
as defining elements $n^r=n^r_{\alpha}\check{e}^{\alpha}$ in $\Lambda^*$.
It is convenient to introduce an auxiliary vector space $V= \mathrm{span}_{\Rh}(e_1,\dots,e_N)$
in which the lattice $\Lambda$ is naturally embedded.

If $a^{\alpha}$ is a globally well defined
$1$-form with values in $V$ then we have variational formulas \eqref{VF} and
\begin{equation}
\mathrm{Hol}_{A^{\alpha}+a^{\alpha}}(\{C_{r},n^{r}\})
=
e^{2\pi i\sum_r\oint_{C_{r}}n^{r}(a)}\,
\mathrm{Hol}_{A^{\alpha}}(\{C_{r},n^{r}\}).
\label{variant2}
\end{equation}
The characteristic property of the holonomy is that if the
contour $C_r$ is a boundary of the disk $D_r$ then
\begin{equation}
\mathrm{Hol}_{A}(\{\pd D_{r},n^{r}\})
=
e^{2\pi i\sum_r\int_{D_{r}}n^{r}(F)}.
\label{variant3}
\end{equation}

Using the variational formulas one obtains the following equations of motion:
\begin{equation}
-\frac{1}{2\pi e^2}\,\lambda^{-1}_{\alpha\beta}\,d*F^{\beta}+
K_{\alpha\beta}F^{\beta}
+ \sum_{r=1}^{N_V} n_{\alpha}^{r}\delta(C_{r})=0
\label{eom}
\end{equation}
where $\delta(C_r)$ is a $2$-form supported on the curve $C_r$.
{}From this equation it follows that
\begin{equation}
K_{\alpha\beta}N^{\beta}+\sum_{r=1}^{N_V} n^r_{\alpha}=0
\label{firstChern}
\end{equation}
where $N^{\alpha}=\int_{\Sigma_g}F^{\alpha}$.
This equation determines the first Chern classes $c_1^{\alpha}$
in the presence of vortices (see details in the beginning of section~\ref{sec:ham:ham}).
This simple arithmetic equation
has solutions iff $\sum_r n^r_{\alpha}\check{e}^{\alpha}$ defines an element of $\Lambda$.
In the FQHE $N^{\alpha}$ has the interpretation of the total number
of electrons in the $\alpha$-th Landau level.

Let $X_3=\Sigma_{g}\times\Rh$ where $\Sigma_{g}$ is
a compact oriented Riemann surface of genus $g$. The metric on $X_3$ is
a product metric:
\begin{equation}
ds^2=-dt^2+ds^2_{\Sigma_g}.
\label{metric}
\end{equation}
In this background we can write $F^{\alpha}=\bar{F}^{\alpha}+dt\wedge
\bar{F}_0^{\alpha}$. Further we will assume that $C_r$ is
a straight line in the time direction passing through the point $x_r$.
Using
\begin{equation}
*_3(dt\wedge\bar{F}_0)=-(*_2\bar{F}_0)
\quad\text{and}\quad
*_3\bar{F}=dt\wedge(*_2\bar{F})
\end{equation}
one can rewrite the classical equations of motion  \eqref{eom} as
\begin{subequations}
\begin{align}
\frac{\delta\mathscr{L}}{\delta\bar{a}}=0:\quad
&\frac{1}{2\pi e^2}\,\lambda^{-1}_{\alpha\beta}
d*\bar{F}^{\beta}+K_{\alpha\beta}\bar{F}_0^{\beta}=-
\frac{1}{2\pi e^2}\,\lambda^{-1}_{\alpha\beta}
\pd_0\bigl(*\bar{F}_0^{\beta}\bigr)
\label{cleom1}
\\
\frac{\delta\mathscr{L}}{\delta\bar{a}_0}=0:\quad
&\frac{1}{2\pi e^2}\,\lambda^{-1}_{\alpha\beta}\,
d*\bar{F}^{\beta}_0+K_{\alpha\beta}\bar{F}^{\beta}+\sum_{r=1}^{N_V} n^r_{\alpha}
\delta^{(2)}(x-x_r)=0.
\end{align}
\label{cleoms}
\end{subequations}

\subsection{Hamiltonian}
\label{sec:ham:ham}
The space of the gauge potentials $\Ac$ is a disjoint union of affine spaces
$\Ac_{\{c_1^{\alpha}\}}$
labeled by the first Chern classes $\{c_1^{\alpha}\}$. The translation group
of the affine space $\Ac_{\{c_1^{\alpha}\}}$ is $\Omega^1(\Sigma_g,V)$.
In the presence
of vortices the gauge potentials $A^{\alpha}$ are connections on nontrivial
line bundles whose first Chern classes $\{N^{\alpha}\}$  satisfy equation \eqref{firstChern}.
In other words $\{A^{\alpha}\}\in\Ac_{\{N^{\alpha}\}}$.
To parameterize the affine space $\Ac_{\{N^{\alpha}\}}$ we choose a reference
potential $A^{\alpha}_{\bullet}$ so an arbitrary gauge potential
$A^{\alpha}=A^{\alpha}_{\bullet}+a^{\alpha}$ where
$a^{\alpha}$ is a globally well defined $1$-form on $X_3$ with values in $V$.
It is convenient to choose the reference potential $A^{\alpha}_{\bullet}$ such that
it satisfies equations \eqref{cleoms} with \textit{only one vortex}
of net charge $\sum_r n^r_{\alpha}$ located on the contour $x_{\bullet}\times\Rh$.

The gauge potential $\{A^{\alpha}\}\in\Ac_{\{N^{\alpha}\}}$ can
be written as
\begin{equation*}
A^{\alpha}=A^{\alpha}_{\bullet}+\bar{a}^{\alpha}+dt\,a_0^{\alpha},\qquad
\bar{f}_0^{\alpha}=\pd_0\bar{a}^{\alpha}-da_0^{\alpha}
\end{equation*}
where $\bar{a}^{\alpha}$ and $a_0^{\alpha}$
are respectively a globally well defined $1$-form and function on $\Sigma_g$ with
values in $V$.
One obtains the following expression for the Lagrangian:
\begin{multline*}
\mathscr{L}=\mathscr{L}_{\bullet}-\frac{1}{2e^2}\int_{\Sigma_g}\lambda^{-1}_{\alpha\beta}\,
d\bar{a}^{\alpha}*d\bar{a}^{\beta}
+\frac{1}{2e^2}\int_{\Sigma_g}\lambda^{-1}_{\alpha\beta}\,
\bar{f}_0^{\alpha}*\bar{f}_0^{\beta}
\\
+\pi\int_{\Sigma_g}\bigl[(d\bar{a},\bar{a}_0)-(\bar{a},\bar{f}_0)\bigr]
+2\pi\sum_{r=1}^{N_V} n^{r}_{\alpha}\bigl[a_0^{\alpha}(x_r)
-a_0^{\alpha}(x_{\bullet})\bigr].
\end{multline*}
Here $\mathscr{L}_{\bullet}$ is the Lagrangian corresponding
to the reference connection $A_{\bullet}^{\alpha}$. The corresponding
action has exactly the form \eqref{action} with $A$ changed to $A_{\bullet}$.
The momentum is defined as
$\delta \mathscr{L}=\int_{\Sigma_g}\Pi_{\alpha}\wedge \delta(\pd_0\bar{a}^{\alpha})$, thus
\begin{equation}
\Pi_{\alpha}=\tilde{\Pi}_{\alpha}-\pi K_{\alpha\beta}\bar{a}^{\beta}
\quad\text{where}\quad
\tilde{\Pi}_{\alpha}=-\frac{1}{e^2}\,
\lambda^{-1}_{\alpha\beta}\,*\bar{f}_0^{\beta}.
\label{PtP}
\end{equation}
The last equation shows that $\tilde{\Pi}_{\alpha}$ is a globally well defined
gauge invariant $1$-form. $\tilde{\Pi}_{\alpha}$ has interpretation
of the electric field.
The symplectic form is
\begin{equation}
\Omega=\int_{\Sigma_g}\delta\Pi_{\alpha}\wedge\delta\bar{a}^{\alpha}
\quad\Rightarrow\quad
\biggl[\int_{\Sigma_g}\phi^{\alpha}\wedge\Pi_{\alpha},\,
a_j^{\beta}(\sigma)\biggr]=i\phi_j^{\beta}(\sigma).
\label{comrel}
\end{equation}

The Hamiltonian is defined by the Legendre transform:
\begin{subequations}
\begin{equation}
\mathscr{L}-\mathscr{L}_{\bullet}=\int_{\Sigma_g}\Pi_{\alpha}\wedge\bar{f}_0^{\alpha}-\mathscr{H}
+\pi\int_{\Sigma_g}(d\bar{a},\bar{a}_0)
+2\pi\sum_{r=1}^{N_V} n^{r}_{\alpha}
\bigl[a_0^{\alpha}(x_r)
-a_0^{\alpha}(x_{\bullet})
\bigr]
\end{equation}
where $\mathscr{H}=\mathscr{H}_m+\mathscr{H}_e$ and
\begin{align}
\mathscr{H}_m&=\frac{1}{2e^2}\lambda^{-1}_{\alpha\beta}\int_{\Sigma_g}
d\bar{a}^{\alpha}*d\bar{a}^{\beta}
\\
\mathscr{H}_e&=\frac{e^2}{2}\,\lambda^{\alpha\beta}
\int_{\Sigma_g}\tilde{\Pi}_{\alpha}*\tilde{\Pi}_{\beta}
.
\label{Helect}
\end{align}
\end{subequations}

\subsection{K\"{a}hler structure on $\Omega^1(\Sigma_g,V)$}
\label{sec:conf.space}
It is important for rest of the paper to notice that
$\mathcal{C}=\Omega^1(\Sigma_g,V)$ has
a natural K\"{a}hler structure.
We will denote by $\mathcal{C}_{\Lambda}=\Omega^1_{\Lambda}(\Sigma_g)$ the subspace of
one forms in $\mathcal{C}$ whose periods lie in $\Lambda$.
The symplectic structure is defined by the bilinear form $K_{\alpha\beta}$
\begin{equation}
\omega_K(\phi_1,\phi_2)=K_{\alpha\beta}\int_{\Sigma_g}\phi_1^{\alpha}\wedge \phi_2^{\beta}.
\label{symplectic}
\end{equation}
Notice that this symplectic form takes integral values on $\mathcal{C}_{\Lambda}$.
The matrix of couplings $\lambda^{\alpha\beta}$ defines a positive definite metric
on $\mathcal{C}$:
\begin{equation*}
\tilde{g}(\phi_1,\phi_2)=\lambda^{-1}_{\alpha\beta}\int_{\Sigma_g}\phi_1^{\alpha}*\phi_2^{\beta}.
\end{equation*}
We will show that the  data of $\omega_K$ and $\tilde{g}$ determine a  complex structure $\mathcal{J}$
on  $\mathcal{C}$. This complex structure is related to the Hodge complex structure by a
matrix $\Gamma$:
\begin{equation}
(\mathcal{J}\cdot \phi)^{\alpha}=\Gamma^{\alpha}{}_{\beta}*\phi^{\beta}.
\label{complexstr}
\end{equation}
$\Gamma$ is a $K$-symmetric matrix ($\Gamma^{tr} K=K\Gamma$) and $\Gamma^2=1$.

In order to extract the matrix $\Gamma$ from the metric $\tilde{g}$ we first
construct the operator $(\tilde{\mathcal{J}}\cdot \phi)^{\alpha}=
\tilde{\Gamma}^{\alpha}{}_{\beta}*\phi^{\beta}$ satisfying
\begin{equation*}
\tilde{g}(\phi_1,\phi_2)=\omega_K(\phi_1,\tilde{\mathcal{J}}\cdot\phi_2).
\end{equation*}
This yields $\tilde{\Gamma}=(\lambda K)^{-1}$. Since $\tilde{\Gamma}$ is $K$-symmetric and $K$ is
a symmetric matrix it follows that the matrix $\tilde{\Gamma}^2$ is positive definite.
Thus there exists a positive definite square root $(\tilde{\Gamma}^2)^{1/2}$.
Now the matrix $\Gamma$ is defined by\footnote{The matrix $\Gamma$
appearing  here is exactly the same matrix $\Gamma$ which was constructed in \cite{Gukov:2004id}.}
\begin{equation}
\Gamma^{\alpha}{}_{\beta}=\bigl(\tilde{\Gamma}(\tilde{\Gamma}^2)^{-1/2}
\bigr)^{\alpha}{}_{\beta}
\quad\text{and}\quad
\tilde{\Gamma}=(\lambda K)^{-1}.
\label{Gamma}
\end{equation}
Notice that if $K$ is a positive definite matrix then $\Gamma=1$.
In general $K$ has indefinite signature and we have the following
statement: $K(1+\Gamma)>0$ and $K(1-\Gamma)<0$. From the construction of $\Gamma$
it follows that $\Gamma\lambda=\lambda\Gamma^{tr}$.

Finally the configuration space $\mathcal{C}$ is actually
a K\"{a}hler vector space with the symplectic structure \eqref{symplectic}
and compatible complex structure \eqref{complexstr}. The metric on $\mathcal{C}$
which is canonically associated to the complex structure $\mathcal{J}$ is
\begin{equation}
g_K(\phi_1,\phi_2):=\omega_K(\phi_1,\mathcal{J}\cdot\phi_2)
=\mu_{\alpha\beta}\int_{\Sigma_g}\phi_1^{\alpha}*\phi_2^{\beta}
\label{g_K}
\end{equation}
where $\mu_{\alpha\beta}=K_{\alpha\gamma}\Gamma^{\gamma}{}_{\beta}$.

\section{Gauss law}
\label{sec:gauss}
\setcounter{equation}{0}
In the following two subsections we derive the quantum Gauss law.
In subsection~\ref{subsec:naive} we derive the Gauss law for the
small gauge transformations using
the Hamiltonian formalism.
In subsection~\ref{subsec:qGL} we derive the quantum
Gauss law using some geometrical facts about (spin) Chern-Simons theory.

\subsection{Naive approach}
\label{subsec:naive}
The classical Gauss law is
\begin{equation}
\mathscr{G}_{\alpha}=\frac{\delta \mathscr{L}}{\delta a_0^{\alpha}}=
-d\bigl[\Pi_{\alpha}-\pi K_{\alpha\beta}\bar{a}^{\beta}\bigr]
+2\pi\sum_{r=1}^{N_V} n^{r}_{\alpha}\bigl[\delta^{(2)}(x-x_r)
-\delta^{(2)}(x-x_{\bullet})\bigr]
=0.
\label{smallGL}
\end{equation}
The \textit{small} gauge transformations $\bar{a}^{\alpha}
\mapsto \bar{a}^{\alpha}+df^{\alpha}$,
where $f^{\alpha}$ are globally well defined, are generated by this Gauss law
\begin{equation}
\psi(a)\stackrel{\text{Gauss law}}{=}
e^{-i\int_{\Sigma_g}f^{\alpha}\mathscr{G}_{\alpha}}
\psi(a^{\alpha})
=
e^{-2\pi i\sum_r n^{r}_{\alpha}[f^{\alpha}(x_r)-f^{\alpha}(x_{\bullet})]}
e^{i\pi \int_{\Sigma_g}(df,\,\bar{a})}
\psi(a+df).
\label{ttaction}
\end{equation}

One can try to generalize the above equation for the \textit{large} gauge
transformations by changing $df\mapsto \omega$ and $f(x)-f(x_{\bullet})\mapsto \int^x_{x_{\bullet}} \omega$
where $\omega$ is a closed $1$-form with periods in the lattice $\Lambda$.
However  this naive procedure fails to give a consistent Gauss law.
This happens because it  ignores the key geometrical
fact that the wave function is a section of a certain line bundle with connection
which has a nonzero curvature. We explain this in detail in the next subsection.

\paragraph{The magnetic translation group.}
Now let us look at the Gauss law \eqref{smallGL}
from a different point of view.
First one sees that the sum of the delta functions in \eqref{smallGL}
is an exact form. Indeed, we can choose a set of non-intersecting contours $\{\Gamma_r\}$
which connect point $x_{\bullet}$ with $\{x_r\}$ and
write $\delta^{(2)}(x-x_r)-\delta^{(2)}(x-x_{\bullet})=d\,\delta(\Gamma_r)$
where $\delta(\Gamma_r)$ is a $1$-form supported on the contour $\Gamma_r$.
Thus the quantity
\begin{equation}
P_{\alpha}=\frac{1}{2\pi}\Pi_{\alpha}-\frac12 K_{\alpha\beta}\bar{a}^{\beta}
-\sum_{r=1}^{N_V}n^r_{\alpha}\delta(\Gamma_r)
\end{equation}
is a closed form. Moreover from the equation of motion \eqref{eom} it
follows that $\pd_0 P=dP_0$ where $P_0$
is a certain function. Thus for any closed time-independent $1$-form $\phi$ with values
in $V$ we can construct a conserved charge
\begin{equation}
P(\phi):=\int_{\Sigma_g}\phi^{\alpha}\wedge P_{\alpha}.
\end{equation}
Although it is not indicated in the notation the charge $P(\phi)$
depends on the choice of contours $\{\Gamma_r\}$.
In section~\ref{subsec:qGL} we will present specific choice
of the contours.

The charges for different $1$-forms $\phi_1$
and $\phi_2$ do not commute:
\begin{equation}
\bigl[P(\phi_1),\,
P(\phi_2)\bigr]
=\frac{i}{2\pi}\,\omega_K(\phi_1,\phi_2)
\end{equation}
where $\omega_K$ is the symplectic form defined in \eqref{symplectic}.
Moreover the charge $P(\phi)$ is not gauge invariant.
To overcome these difficulties we have to choose a polarization
on $\mathcal{C}$.
So, choose a ``canonical'' homology basis $\mathscr{A}=\{a_p,b^p\}$ on $\Sigma_g$,
$p=1,\dots,g$.
It defines a dual basis $\mathscr{A}^*=(\alpha^p,\beta_p)$
in the cohomology.
Then we can write $\phi=\phi_a+\phi_b$
where $\phi_a$ ($\phi_b$)
is a part of $\phi$ which contains only $\alpha^{p}$ ($\beta_p$).
If $\phi$ has periods in $\Lambda^*$
then the following operators are gauge invariant:
\begin{equation}
\mathtt{W}_a(\phi)=e^{-2\pi iP(\phi_a)}
\quad\text{and}\quad
\mathtt{W}_b(\phi)=e^{-2\pi iP(\phi_b)}
\label{W}
.
\end{equation}
These operators satisfy the following commutation relations:
\begin{equation}
\mathtt{W}_a(\phi)\mathtt{W}_b(\phi)
=e^{-2\pi i
\omega_K(\phi_a,\phi_b)}
\,\mathtt{W}_b(\phi)\mathtt{W}_a(\phi).
\end{equation}
Notice that if $\phi\in H^1(\Sigma_g,\Lambda)$
then $\mathtt{W}(\phi)$ is a gauge transformation.
Thus we can identify $\phi$ with $\phi+\xi$
where $\xi$ is any vector in $H^1(\Sigma_g,
\Lambda)$.
With this identification $\mathtt{W}_a$ and $\mathtt{W}_b$
generate the finite Heisenberg group $\mathcal{W}$:
\begin{equation}
0\to (\Lambda^*/\Lambda)^{\otimes g}\to \mathcal{W}\to
H^{1}(\Sigma_g,\Lambda^*/\Lambda)\to 0.
\end{equation}
This group is also known as the magnetic translation group.
The Hilbert space of the theory should be
a representation of this group; we confirm this
below.

\subsection{Quantum Gauss Law}
\label{subsec:qGL}
The classical Gauss law fixes the first Chern class of the connection on $P$ (see eq.~\ref{firstChern}).
The space of allowed connections is then an affine space
modelled on $\mathcal{C}=\Omega^1(\Sigma_g,V)$.
So, in the presence of vortices the configuration space is
\begin{equation*}
\mathscr{C}=\mathcal{C}\times \Sigma_g^{N_V}.
\end{equation*}
Here the second factor describes the position of the vortices.
When evaluated on a manifold with boundary the spin Chern-Simons action
is a section of a line bundle $\mathcal{L}_{CS}$ over $\mathcal{C}$.
It was explained at the end of section~\ref{sec:aCS} that this line bundle
depends on a choice of spin structure $\sigma$.
Thus the quantum wave function is a section of
\begin{equation}
\mathcal{L}_{\sigma}=\mathcal{L}_{CS}\otimes E_{n^1}\otimes
\dots\otimes E_{n^{N_V}}\to \mathscr{C}
\end{equation}
where $E_{n}\to \Sigma_g$ is the line bundle associated to the principal
bundle $P$ in which the group $U(1)^N$ acts in the representation
$n_{\alpha}$.
The line bundle $\mathcal{L}_{\sigma}$ has a natural connection
defined by the spin Chern-Simons phase \eqref{CSphase}
together with the holonomies.
Consider a path $p(t)=\{A^{\alpha}(t)\}\times\{x_r(t)\}$
in the space of connections $\Ac_{N^{\alpha}}$ and $\Sigma_g$, $t\in[0,1]$ is
the coordinate on the path.
Then the parallel
transport in $\mathcal{L}_{\sigma}$ is defined by:
\begin{equation}
\mathscr{U}(p) :=e^{2\pi i \CS^{\text{spin}}_{\Sigma_g\times[0,1]}(A^{\alpha}(t))}
\,\mathrm{Hol}_{A^{\alpha}(t)}(\{x_r(t),n^{r}\})\in \mathrm{Hom}
\bigl(\mathcal{L}_{\sigma}\bigl|_{\{A^{\alpha}(0),x_r(0)\}},
\mathcal{L}_{\sigma}\bigl|_{\{A^{\alpha}(1),x_r(1)\}}\bigr).
\label{connection}
\end{equation}
\begin{figure}[!t]
\centering
\includegraphics[width=380pt]{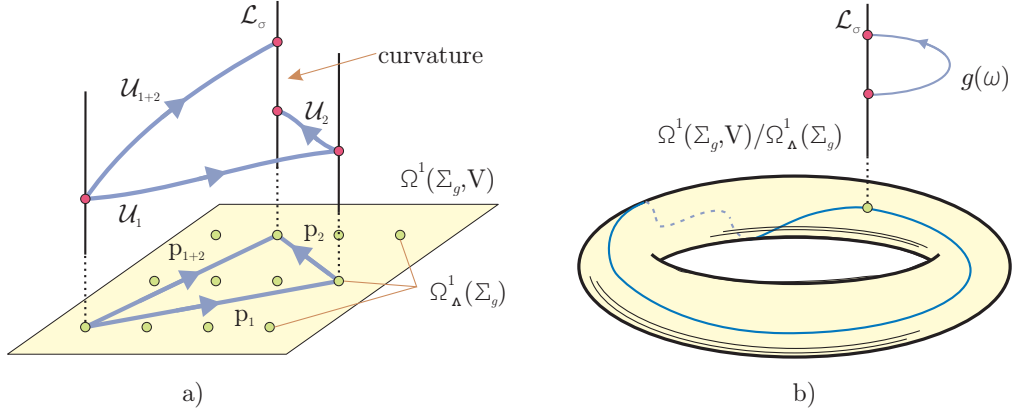}
\caption{Figure a) shows that the line bundle
$\mathcal{L}_{\sigma}$ has non-zero curvature, and
thus the parallel transport does not define the lift of
the gauge group. In figure b) we present the
gauge invariant configuration space $\mathcal{C}$. The parallel transport
along straight line path on figure a) corresponds to the holonomy around the closed
loop in $\mathcal{C}/\mathcal{C}_{\Lambda}$.}
\label{fig:tor}
\end{figure}
The tangent vector to the path $p$ is a pair $(\phi^{\alpha},\,\{\xi_r\})$
where $\phi\in\Omega^1(\Sigma_g,V)$ and $\xi_r\in T_{x_r(0)}\Sigma_g$
is a tangent vector at the point $x_r=x_r(0)$.
The curvature of the connection \eqref{connection} is
\begin{equation}
\Omega\bigl((\phi_1,\{\xi_r^{(1)}\}),\,(\phi_2,\{\xi_r^{(2)}\})\bigr)
=-\omega_K(\phi_1,\,\phi_2)
+\sum_{r=1}^{N_V}n^{r}_{\alpha}F^{\alpha}(x_r)(\xi^{(1)}_r,\,\xi^{(2)}_r).
\label{Omegacurv}
\end{equation}
Now, for any 1-form $\phi^{\alpha}$  introduce the straightline path
in the space of connections $\Ac_{N^{\alpha}}$
\begin{equation}
p_{A;\,\phi}(t)=\{A^{\alpha}+t\phi^{\alpha}\},\qquad t\in[0,1].
\label{path}
\end{equation}
Here we do not move the position of the vortices.
Using the formula for the curvature   we find
\begin{equation}
\mathscr{U}(p_{A;\,\phi_1+\phi_2}\times\{x_r\})
=
\mathscr{U}(p_{A+\phi_1;\,\phi_2}\times\{x_r\})
\,\mathscr{U}(p_{A;\,\phi_1}\times\{x_r\})\,
e^{+i\pi
\omega_K(\phi_1,\phi_2)
}.
\label{Umult}
\end{equation}

It follows from \eqref{Umult} that  parallel transport
\textit{does not} define a lift
of the gauge group to the total space of $\mathcal{L}_{\sigma}$ (see Figure~\ref{fig:tor}).
To define the lift of the
group action we choose a standard path, say \eqref{path},
and define the action on a section $\Psi_{\sigma}$ of $\mathcal{L}_{\sigma}$ by
\begin{equation}
\bigl(g(\omega)\cdot\Psi_{\sigma}\bigr)(A+\omega,x_r)=
\tilde{\varphi}_{\sigma}^*(A,\,\omega;\{x_r\})\,
\mathscr{U}(p_{A;\,\omega}\times\{x_r\})\,
\Psi_{\sigma}(A,x_r)
\label{gact1}
\end{equation}
where $\tilde{\varphi}_{\sigma}(A,\omega;\{x_r\})$ is a phase, and $\omega$
is an element of $\mathcal{C}_{\Lambda}$
(a closed one form with periods in $\Lambda$). The  ``lifting phase''
 $\tilde{\varphi}_{\sigma}(A,\omega;\{x_r\})$ must satisfy
\begin{equation}
\tilde{\varphi}_{\sigma}(A,\omega_1+\omega_2;\{x_r\})
=\tilde{\varphi}_{\sigma}(A+\omega_1,\omega_2;\{x_r\})
\tilde{\varphi}_{\sigma}(A,\omega_1;\{x_r\})
\,e^{+i\pi \omega_K(\omega_1,\omega_2)}.
\label{liftingrel}
\end{equation}

\subsubsection{The cocycle}
This subsection is a little technical.
First, we construct the canonical cocycle $\varphi_{\sigma}(A,\omega; \{x_r\})$,
see equation \eqref{varphi}. However it happens that
this is not the cocycle we need to construct the lift of the gauge group.
Other cocycles differ from the canonical one by the local
terms. Second, we fix this local term.
The final result is summarized by equations \eqref{theCocycle}
and \eqref{theSpin}.

\begin{figure}[t]
\centering
\includegraphics[width=340pt]{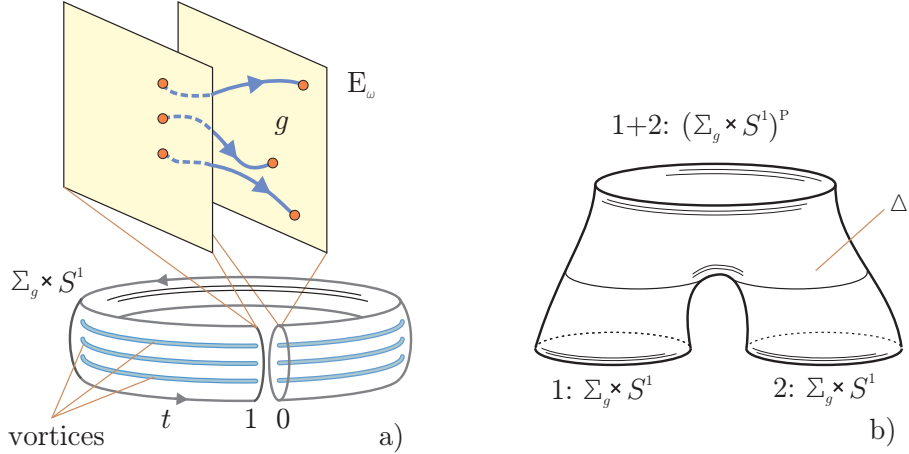}
\caption{In figure a) we illustrate the construction of the twisted
vector bundle $E_{\omega}$ over $\Sigma_g\times S^1_-$. The twist is defined by
identification of $E|_{t=1}$ with $E|_{t=0}$ by the gauge transformation
$g$ corresponding to $\omega$.
In figure b) we present the surface $\Delta$ (a pair of pants)
which is used to prove the cocycle relation.}
\label{fig:cocycle}
\end{figure}

\paragraph{Construction.}
The cocycle can be constructed using the phase \eqref{CSphase} \cite{Witten:1996hc}.
To construct the cocycle we proceed as in \cite{Diaconescu:2003bm}.
Consider the following three dimensional
manifold $\Sigma_g\times S^1_-$ where we choose the
antiperiodic spin structure on the circle. (Only the antiperiodic spin structure
on the circle can be extended to a spin structure on the disk).
Construct the twisted bundle $E_{\omega}$ over $\Sigma_g\times S^1_-$ (see Figure~\ref{fig:cocycle}a):
\begin{equation}
E_{\omega}=E\times[0,1]/\{(p,0)\sim (g\cdot p,1)\}
\end{equation}
where $g$ corresponds to a gauge transformation $\omega$, and
$E$ was defined in the beginning of section~\ref{sec:spinCS}.
The connection  satisfies $A(1)=A(0)+\omega$. The cocycle is
the spin Chern-Simons phase on this $3$-fold times holonomies
\begin{equation}
\varphi_{\sigma}(A,\omega;\{x_r\})
=e^{2\pi  i\CS^{\text{spin}}_{\Sigma_g\times S^1_-}(A+t\omega)}
\mathrm{Hol}_{A+t\omega}(\{x_r\times S^1_-;n^r\}).
\label{varphi}
\end{equation}
One sees that the connection $A+t\omega$ does not have indices along the $S^1$ direction
which means that the term with the holonomy is a locally constant function of $A$.
$\varphi_{\sigma}$ defined in this way satisfies the cocycle relation \eqref{liftingrel}.
The proof is a standard cobordism
argument. We consider a connection
on the $3$-manifold $(\Sigma_g\times S^1_-)\cup(\Sigma_g\times S^1_-)\cup(\Sigma_g\times S^1_-)^P$
restricting to $A^{\alpha}+t\omega_1^{\alpha},\,
A^{\alpha}+\omega_1^{\alpha}+t\omega_2^{\alpha},\,
A+t(\omega_1^{\alpha}+\omega_2^{\alpha})$ on the three
components ($X^P$ means change of orientation of manifold $X$).
Then we choose the extending spin $4$-manifold
to be $Z_4 = \Sigma_g\times \Delta$ where $\Delta$ is
a pair of pants bounding the three circles with spin structure restricting to
$S^1_-$
 on the three components (see Figure~\ref{fig:cocycle}b).
To be explicit we can choose $\Delta$ to be the simplex $\{(t_1, t_2) : 0 \leqslant t_2
\leqslant t_1 \leqslant 1\}$
with identifications $t_i \sim t_i + 1$.
Now we extend the connection to be $\tilde{A}(t_1,t_2)=A+t_1\omega_1+t_2\omega_2$
which clearly restricts to the required connection on the boundary.
The field strength is $F_Z=F+dt_1\wedge\omega_1+dt_2\wedge\omega_2$.
 We can therefore use \eqref{CSphase} to show that the product of phases around the
boundary is
\begin{equation*}
\varphi_{\sigma}(A,\omega_1;\{x_r\})\varphi_{\sigma}(A+\omega_1,\omega_2;
\{x_r\})\varphi^*_{\sigma}(A,\omega_1+\omega_2;\{x_r\})
=e^{i\pi K_{\alpha\beta}\int_{\Sigma_g\times\Delta}F_Z^{\alpha}\wedge F_Z^{\beta}}
=e^{-i\pi \omega_K(\omega_1,\omega_2)}.
\end{equation*}
This proves the desired cocycle relation.

The dependence of the cocycle \eqref{varphi} on $A$ and the position of the vortices
$\{x_r\}$ can
be extracted from the variational formulas \eqref{variat},
\eqref{variant2} and \eqref{variant3}:
\begin{subequations}
\begin{align}
\varphi_{\sigma}(A+a,\omega;\{x_r\})&=
e^{2\pi i\,\omega_K(\omega,a)}\varphi_{\sigma}(A,\omega;\{x_r\})
\\
\varphi_{\sigma}(A,\omega;\{x_r+C_r\})&=
e^{-2\pi i\,\sum_r \int_{C_r}n^r(\omega)}\,\varphi_{\sigma}(A,\omega;\{x_r\})
\end{align}
\label{var-varphi}
where $a$ is an arbitrary element of $\Omega^1(\Sigma_g,V)$
and $\{C_r\}$ is an arbitrary set of nonintersecting contours starting at
points $\{x_r\}$.
\end{subequations}


We expect that there are $2^{2g}$ different cocycles
corresponding to $2^{2g}$ different spin structures (see Figure~\ref{fig:picard}).
The group $H^1(\Sigma_g,\frac12\Zh/\Zh)$ acts transitively on the
space of spin structures.
Let $\epsilon\in H^1(\Sigma_g,\frac12\Zh/\Zh)$
then from equation \eqref{VF:spin} it follows that the cocycles
corresponding to spin structures $\sigma$ and $\sigma+\epsilon$
are related by
\begin{equation*}
\varphi_{\sigma+\epsilon}(A,\omega;\{x_r\})=\varphi_{\sigma}(A,\omega;\{x_r\})\,
e^{2 \pi i \int_{\Sigma_g\times S^1_-}W_2(F)\wedge\epsilon}
\end{equation*}
where $W_2$ is \textit{any} representative of the characteristic class
$[W_2]\in \Lambda^*/2\Lambda^*$
which was defined in section~2.3.
Recall that $F=F_A+dt\wedge \omega$ and thus we have
\begin{equation}
\varphi_{\sigma+\epsilon}(A,\omega;\{x_r\})=\varphi_{\sigma}(A,\omega;\{x_r\})\,
e^{2\pi i\,\omega_K(\omega,\,\epsilon\otimes W_2)}.
\label{varphi_spin}
\end{equation}

\paragraph{Dependence on the base point.}
The cocycle \eqref{varphi} is not unique. The are other cocycles which
differ by local terms.
In section~\ref{sec:ham:ham} we
chose a very particular coordinate system in the space of connections
and vortices.
Using the variational formulas \eqref{var-varphi} we
can the express the cocycle $\varphi_{\sigma}(A,\omega;\{x_r\})$
through its value at the base point $A_{\bullet},\,x_{\bullet}$:
\begin{equation}
\varphi_{\sigma}(A_{\bullet}+a,\omega;\{x_{\bullet}+\Gamma_r\})
=\exp\Bigl[-2\pi i \sum_{r=1}^{N_V}\int_{\Gamma_r}n^r(\omega)\Bigr]\,
e^{2\pi i\omega_K(\omega,a)}\,
\varphi_{\sigma}(A_{\bullet},\omega; x_{\bullet})
\label{varphi2}
\end{equation}
where $\{\Gamma_r\}$ is a set of nonintersecting contours
connecting $x_{\bullet}$ with $\{x_r\}$. A particular choice
of these contours will be explained in the beginning of the next section.
Notice however that the cocycle \eqref{varphi2} does not
depend on the homotopy class of the contours.
The cocycle at the base point satisfies
\begin{equation}
\varphi_{\sigma}(A_{\bullet},\omega_1+\omega_2;x_{\bullet})
=\varphi_{\sigma}(A_{\bullet},\omega_2;x_{\bullet})
\varphi_{\sigma}(A_{\bullet},\omega_1;x_{\bullet})
\,e^{i\pi\omega_K(\omega_1,\omega_2)}.
\end{equation}
{}From this equation it follows that $\varphi_{\sigma}(A_{\bullet},\omega)$
is \textit{linear} on the space of closed $1$-forms with periods in $2\Lambda$:
\begin{equation}
\varphi_{\sigma}(A_{\bullet},\omega;x_{\bullet})\Bigr|_{\omega\in\mathcal{C}_{2\Lambda}}=
e^{2\pi i \omega_K(\omega,\tilde{\Xi}_{\bullet})}.
\label{1}
\end{equation}
This equation determines $\tilde{\Xi}$ modulo $\Omega^1_{\frac12\Lambda}(\Sigma_g)$.
On the other hand using \eqref{variant3} we can explicitly evaluate
the cocycle for a small gauge transformation:
\begin{equation}
\varphi_{\sigma}(A_{\bullet},df;x_{\bullet})=e^{-2\pi iK_{\alpha\beta}\int_{\Sigma_g}f^{\alpha} F_{\bullet}^{\beta}
-2\pi i \sum_{r}n^r_{\alpha}f^{\alpha}(x_{\bullet})}.
\label{2}
\end{equation}
Recall that $F_{\bullet}$ satisfies the equation of motion \eqref{eom}
and thus the combination appearing in the exponential is an exact form:
\begin{equation}
K_{\alpha\beta}F^{\beta}_{\bullet}
+ \sum_{r=1}^{N_V} n_{\alpha}^{r}\delta(x_{\bullet})=
\frac{1}{2\pi e^2}\,\lambda^{-1}_{\alpha\beta}\,d*F_{\bullet,0}^{\beta}
\equiv d\Xi_{\bullet}.
\label{eom1}
\end{equation}
The consistency of \eqref{1} with \eqref{2} requires that
$d(\tilde{\Xi}_{\bullet}-\Xi_{\bullet})=0$. Thus $\tilde{\Xi}_{\bullet}$
is just another trivialization of the left hand side of \eqref{eom1}.

Recall that $\tilde{\Xi}_{\bullet}$
is determined by equation \eqref{1}
only modulo $\Omega^1_{\frac12\Lambda}(\Sigma_g)$.
To define the cocycle we must fix this ambiguity
modulo $\Omega^1_{\Lambda}(\Sigma_g)$.
We claim that there is a distinguished choice of $\tilde{\Xi}_{\bullet}\mod \mathcal{C}_{\Lambda}$.
To prove this we
choose a solution $\tilde{\Xi}_{\bullet}$ of $\eqref{1}$. Then
an arbitrary solution is of the form $\tilde{\Xi}_{\bullet}+\frac12\eta$
where $\eta\in\mathcal{C}_{\Lambda}$.
Now we introduce the cocycle
\begin{equation}
f_{\sigma}^{\bullet}(\omega)=
\varphi_{\sigma}^{*}(A_{\bullet},\omega;x_{\bullet})\,
e^{2\pi i\omega_K(\omega,\tilde{\Xi}_{\bullet})+i\pi\omega_K(\omega,\eta)}.
\label{fbullet}
\end{equation}
First, notice that the function $f^{\bullet}_{\sigma}(\omega)$
 descends to $\mathcal{C}_{\Lambda}/\mathcal{C}_{2\Lambda}$
and satisfies the cocycle law.
Second, note since $f^{\bullet}_{\sigma}(\omega)^2=f^{\bullet}_{\sigma}(2\omega)=1$
it follows that $f^{\bullet}_{\sigma}(\omega)=\pm 1$.
Now a choice of a canonical basis $\mathscr{A}^{*}=\{\alpha^p,\beta_p\}$ in $H^1(\Sigma_g,\Zh)$
determines a preferred spin structure $\sigma_{\mathscr{A}}$
(see above equation \eqref{mod2index}). In the coordinates,
$\omega=m^1_p\alpha^p+m_2^p\beta_p$, the cocycle
$f^{\bullet}_{\sigma_{\mathscr{A}}}(\omega)$ is necessarily of the form
\begin{equation*}
f^{\bullet}_{\sigma_{\mathscr{A}}}(\omega)=
\exp\Bigl\{
i\pi (m^1_p,m_2^p)
+i\pi[(m^1_p,x^p_2)-(m_2^p,x_p^1)]
\Bigr\}
\end{equation*}
where $x^1_p,x^p_2\in\Lambda$.
Now the distinguished choice of $\eta$ is determined by
setting $x^1=x_2=0$, so $f^{\bullet}_{\sigma_{\mathscr{A}}}(\omega)=e^{i\pi (m^1_p,m_2^p)}$.
Having done this the variational formula determines
\begin{equation*}
f^{\bullet}_{\sigma_{\mathscr{A}}+\epsilon}(\omega)=
\exp\Bigl\{
i\pi (m^1_p,m_2^p)
+2\pi i\bigl[\epsilon^1_p\otimes W_2(m_2^p)
-\epsilon_2^p\otimes W_2(m^1_p)\bigr]
\Bigr\}
\label{fbullet:coord}
\end{equation*}
Using the affine transformation
of the characteristics $(\epsilon^1,\epsilon_2)$
one can show that this choice is in fact $Sp(2g,\Zh)$ invariant.
Hence $\tilde{\Xi}_{\bullet}\mod \mathcal{C}_{\Lambda}$
determined in this way is independent of choice of basis in $H^1(\Sigma_g,\Zh)$.

\paragraph{Summary.} The cocycle $\tilde{\varphi}_{\sigma}$ which we use to define
the lift of the gauge group \eqref{gact1} is
\begin{equation}
\tilde{\varphi}^*_{\sigma}(A,\omega;\{x_r\})
=\varphi_{\sigma}^{*}(A_{\bullet}+a,\omega;\{x_{\bullet}+\Gamma_r\})\,
e^{2\pi i\omega_K(\omega,\tilde{\Xi}_{\bullet}+\frac12\eta)}
=f_{\sigma}^{\bullet}(\omega)\,e^{-2\pi i\omega_K(\omega,a)
+2\pi i \sum_{r=1}^{N_V}\int_{\Gamma_r}n^r(\omega)}
\label{theCocycle}
\end{equation}
where $f^{\bullet}_{\sigma}(\omega)$ is defined by \eqref{fbullet}.
It is a $\Zh_2$-valued function on $\mathcal{C}_{\Lambda}/\mathcal{C}_{2\Lambda}$:
$f^{\bullet}_{\sigma}(df^{\alpha})=1$, $f^{\bullet}_{\sigma}(2\omega)=1$ and $f^{\bullet}_{\sigma}(\omega)=\pm 1$.
The dependence on the spin structure is summarized by the following
variational formula
\begin{equation}
f_{\sigma+\epsilon}^{\bullet}(\omega)=
e^{-2\pi i\, \omega_K(\omega,\epsilon\otimes W_2)}\,f_{\sigma}^{\bullet}(\omega).
\label{theSpin}
\end{equation}

\subsubsection{The gauge transformations}
\label{sec:gtr}
\begin{figure}[!t]
\centering
\includegraphics[width=380pt]{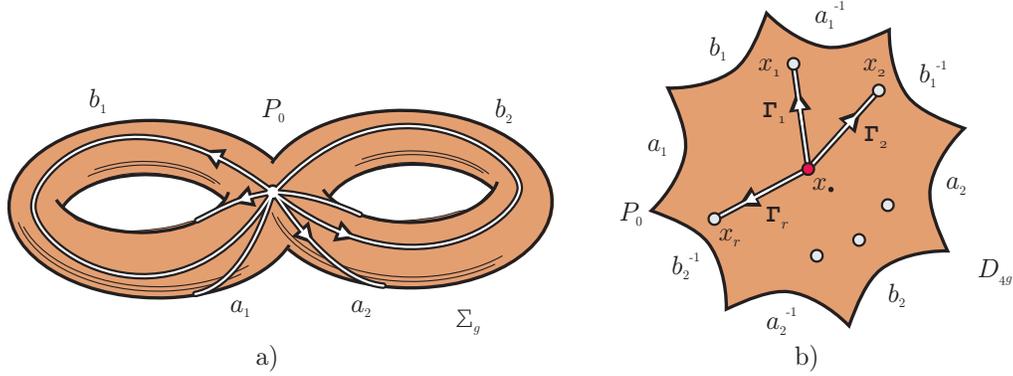}
\caption{In Figure a) we present genus two Riemann surface with
marking. In Figure b) we use the $a$ and $b$ cycles to cut the Riemann
surface $\Sigma_g$ to obtain a connected  $4g$-gon.
The straight white lines represent the canonical paths which
were used to trivialize the line bundle $\mathcal{L}_{\sigma}.$}
\label{fig:cuts}
\end{figure}

To define the action of the gauge group on the wave functions,
we must first trivialize the line bundle $\mathcal{L}_{\sigma}$.
We have already trivialized the space of connections $\Ac_{\{N^{\alpha}\}}$
by choosing a reference connection $A_{\bullet}$ in section~\ref{sec:ham:ham}.
We must also trivialize the space of the vortex configurations on
$\Sigma_g$. To this end we endow $\Sigma_g$ with marking $\mathscr{A}$,
i.e. we choose a point $P_0$ and a ``canonical'' basis of one cycles
$\{a_p,b^p\}$, $p=1,\dots,g$ (see Figure~\ref{fig:cuts}a).
Next we cut the Riemann surface
along these cycles and obtain a $4g$-gon $D_{4g}$ presented in Figure~\ref{fig:cuts}b.
Then the vortices $x_r$ can be canonically connected to $x_{\bullet}$ by a
set of straight line contours $\Gamma_r$ (see section~\ref{sec:ham:ham} for definition of $x_{\bullet}$).
The space $\tilde{\mathscr{C}}=\Omega^1(\Sigma_g,V)\times D_{4g}^{N_V}$ is contractible and thus
the line bundle $\mathcal{L}_{\sigma}$ over $\tilde{\mathscr{C}}$ is trivial.
Using first the parallel transport of the vortices along the straight line pathes
$\{\Gamma_r\}$, and then
the  parallel transport of the gauge field along the path \eqref{path} keeping the vortices fixed
we define a canonical nowhere vanishing
section $S(A,x_r)$ of $\mathcal{L}_{\sigma}$. More precisely,
\begin{equation*}
S(A_{\bullet}+\bar{a},x_{\bullet}+\Gamma_r)=
\mathscr{U}(p_{A_{\bullet},\bar{a}}\times\{x_{r}\})\,
\mathscr{U}(\{A_{\bullet}\}\times\{\Gamma_r\})\,S_{\bullet}.
\end{equation*}
The {\it ratio} $\psi_{\sigma}(\bar{a},x_r) :=  \Psi_{\sigma}(A_{\bullet}+\bar{a},x_r)/S(A_{\bullet}
+\bar{a},x_r)$
is   a function, rather than a section
(to simplify the notations we will write $\bar{a}$
instead of $A_{\bullet}+\bar{a}$ in the argument of the wave function). The action of the   gauge group on this
function is
\begin{equation}
\bigl(\mathfrak{g}(\omega)\cdot\psi_{\sigma}\bigr)(\bar{a}+\omega,x_r)
=\tilde{\varphi}_{\sigma}^*(A_{\bullet}+\bar{a};\omega)\,
e^{-i\pi\omega_K(\bar{a},\omega)}
\psi_{\sigma}(\bar{a},x_r)
\label{largegauge}
\end{equation}
where the cocycle $\tilde{\varphi}_{\sigma}^{*}$ is defined by \eqref{theCocycle}.
The Gauss law $g\cdot\Psi(A^{\alpha})=\Psi(g\cdot A^{\alpha})$
for the wave function takes the form
\begin{equation}
\psi_{\sigma}(\bar{a}+\omega,x_r)=
f^{\bullet}_{\sigma}(\omega)\,
e^{i\pi\omega_K(\bar{a},\omega)+2\pi i\sum_r\int_{\Gamma_r}n^{r}_{\alpha}\omega^{\alpha}}
\psi_{\sigma}(\bar{a},x_r)
\label{Gausslaw}
\end{equation}
Since for the small gauge transformations $f^{\bullet}_{\sigma}(df^{\alpha})=1$
this Gauss law for $\omega^{\alpha}=df^{\alpha}$ agrees with the Gauss law \eqref{ttaction}.
This fact partially explains the statement that the wave function is a section of $\mathcal{L}_{\sigma}$.

{}From the geometrical point of view the Gauss law
\eqref{Gausslaw} says that the gauge invariant wave function is
a section of a line bundle $\mathcal{L}_{\sigma}$ over $\mathscr{C}/\mathcal{C}_{\Lambda}$.
The symplectic form $\omega_K$
descends to a nontrivial element of $H^2(\mathcal{C}/\mathcal{C}_\Lambda,\Zh)$.
{}From \eqref{Omegacurv} it follows that the first Chern class of $\mathcal{L}_{\sigma}$
along $\mathcal{C}$
is $[-\omega_K]$.


\section{Physical wave function}
\label{sec:wave}
\setcounter{equation}{0}
Using the Hodge decomposition we can write $\bar{a}^{\alpha}$ as
\begin{equation}
\bar{a}^{\alpha}=a_h^{\alpha}+a'{}^{\alpha}+a''{}^{\alpha}
\end{equation}
where $a_h\in\mathrm{Harm}^1(\Sigma_g,V)$, $a'$ is the projection
on the image of $d^{\dag}=*d*$ and $a''$ is the projection on the image of $d$.
This is an orthogonal decomposition with respect to the
Hodge metric on $p$-forms. The magnetic Hamiltonian $\mathscr{H}_m$
depends only on $a'$, while the electric Hamiltonian $\mathscr{H}_e$
is a sum of terms depending on $a_h$, $a'$ and $a''$. Thus the
wave function $\Psi_{\sigma}$ factorizes as a product $\Psi_{\text{harm}}\Psi_{\text{massive}}$.
The massive sector has a unique groundstate, but the harmonic sector
has many ground states, leading to a Hilbert space of ``conformal blocks''.
We will be mostly focussing on the harmonic sector.

\subsection{Basis}
\label{sec:basis}
Recall that the configuration space $\mathcal{C}$ has a natural
complex structure $\Jc$ defined by eq.~\eqref{complexstr}. $\Jc$
consists of two operators: the Hodge $*$ and matrix $\Gamma$.
Recall that $\Gamma^2=1$.
Using $*$ we can decompose $\Omega^1(\Sigma_g,V)=
\Omega^{1,0}(\Sigma_g,V)\oplus \Omega^{0,1}(\Sigma_g,V)$.
Notice that $\Omega^{1,0}(\Sigma_g,V)$ are holomorphic forms
with respect to the Hodge complex structure which in the case
of a Riemann surface is opposite to the standard complex structure.
In other words $\Omega^{0,1}$ corresponds to abelian differentials $\sim dz$
while $\Omega^{1,0}$ to abelian antidifferentials $\sim d\oz$.
Using $\Gamma$ we can construct two projection operators
$P_{\pm}=\frac12(1\pm\Gamma)$ and decompose $V=V_+\oplus
V_-$. Thus with respect to the complex structure $\Jc$ we have the following
decomposition of $\mathcal{C}$:
\begin{equation}
\mathcal{C}=\mathcal{C}^{1,0}\oplus\mathcal{C}^{0,1}
=\Bigl[\Omega^{1,0}(\Sigma_g,V_+)\oplus\Omega^{0,1}(\Sigma_g,V_-)\Bigr]
\oplus\Bigl[\Omega^{1,0}(\Sigma_g,V_-)\oplus\Omega^{0,1}(\Sigma_g,V_+)\Bigr].
\label{Cdecom}
\end{equation}
Now we have to choose a basis in this decomposition.

For the Riemann surface there is a natural way to choose a basis
of harmonic $1$-forms.
We choose a ``canonical'' homology basis $\mathscr{A}=\{a^p,b_p\}$
($p=1,\dots,g$)  in $H_1(\Sigma,\Zh)$.
The homology basis defines a dual basis of $1$-forms
$\mathscr{A}^*=\{\alpha^p,\beta_p\}$.

It is convenient to normalize
a basis $\{\zeta^p\}$ of abelian differentials
 such that $\zeta^p$ has period
one through the cycle $a^p$ and zero through the cycles $a^q$, $q\ne p$.
Then the integrals over the $b$-cycles are fixed and define the period matrix:
\begin{equation}
\oint_{a^p}\zeta^q=\delta^q{}_p
\quad\text{and}\quad
\oint_{b_p}\zeta^q=\tau^{qp}_{\mathscr{A}}.
\label{zeta_basis}
\end{equation}
The subscript $\mathscr{A}$ shows that the period matrix
depends on a choice of the symplectic basis (in some formulas
below we will omit it).
Decomposing into real and imaginary parts, $\tau_{\mathscr{A}}=\tau_1+i\tau_2$
is a symmetric matrix with
positive definite imaginary part $\tau_2$.
Relations
\eqref{zeta_basis} imply
that $\zeta^p=\alpha^p+\tau^{pq}\beta_q$.
Since the Hodge complex structure is opposite to the usual one we have the relations
\begin{equation}
\int_{\Sigma_g}\bar{\zeta}^p\wedge \zeta^q
=2i\, \tau_2^{pq}
\quad\text{and}\quad
\int_{\Sigma_g}\bar{\zeta}^p *\zeta^q=2\, \tau_2^{pq}.
\label{zz}
\end{equation}
It is easy to see that an arbitrary harmonic 1-form with values in $V$:
$C=C_p^{1\,\alpha}\,\alpha^p\otimes e_{\alpha}+
C^{p\,\alpha}_2\,\beta_p\otimes e_{\alpha}$
decomposes under \eqref{Cdecom} as $C=C^{1,0}+C^{0,1}$ where
\begin{equation}
C^{1,0}=-\frac{1}{2i}(C_2-C^1\cdot\tau)_+^{\alpha}\cdot\tau_2^{-1}\cdot\bar{\zeta}\otimes e_{\alpha}
+\frac{1}{2i}(C_2-C^1\cdot\bar{\tau})_-^{\alpha}\cdot\tau_2^{-1}\cdot\zeta\otimes e_{\alpha}
\quad\text{and}\quad
C^{0,1}=\overline{C^{1,0}}.
\label{holsplit}
\end{equation}
The index $\pm$ denotes the projection by $P_{\pm}$.

\subsection{Hamiltonian equation}
We can expand the gauge fields $a^{\alpha}$ and momenta $\Pi_{\alpha}$ in the
basis $\mathscr{A}^*$:
\begin{equation}
a^{\alpha}=a^{1\,\alpha}_p\,\alpha^p+a_2^{p\,\alpha}\beta_p
\quad\text{and}\quad
\Pi_{\alpha}=\Pi^{2}_{p\,\alpha}\,\alpha^p-\Pi_{1\,\alpha}^{p}\beta_p.
\end{equation}
The nonzero commutation relations \eqref{comrel} are
\begin{equation}
[\Pi_{1\,\alpha}^p,\,a^{1\,\beta}_q]=-i\delta^{p}{}_q\,\delta^{\beta}{}_\alpha
\quad\text{and}\quad
[\Pi_{\alpha}^{2\,p},\,a^{\beta}_{2\,q}]=-i\delta^{p}{}_q\,\delta^{\beta}{}_\alpha.
\end{equation}

Certainly it is more convenient to use holomorphic coordinates,
so we introduce
\begin{equation}
a_{\tau}=a_2-\tau\cdot a^1
\quad\text{and}\quad
\Pi^{\tau}=-\frac{1}{2i}\,\tau_2^{-1}\cdot(\Pi_1+\bar{\tau}\cdot\Pi^2)
\end{equation}
and similarly for $a_{\bar{\tau}}$ and $\Pi^{\bar{\tau}}$.
This yields the following nonzero commutation relations:
\begin{equation}
[\Pi^{\tau}_{\alpha, p},\,a^{\beta,q}_{\tau}]=-i\delta_{p}{}^{q}\delta_{\alpha}{}^{\beta}
\quad\text{and}\quad
[\Pi^{\bar{\tau}}_{\alpha,p},\,a^{\beta,q}_{\bar{\tau}}]=-i\delta_{p}{}^{q}\delta_{\alpha}{}^{\beta}.
\end{equation}

The Hamiltonian \eqref{Helect} in the harmonic sector is
\begin{equation}
\mathscr{H}_{\text{harm}}
=e^2\,\lambda^{\alpha\beta}
\bigl[\tilde{\Pi}_{\alpha}^{\tau}\cdot\tau_2\cdot\tilde{\Pi}^{\bar{\tau}}_{\beta}
+\tilde{\Pi}_{\alpha}^{\bar{\tau}}\cdot\tau_2\cdot\tilde{\Pi}^{\tau}_{\beta}
\bigr]
\end{equation}
where
\begin{equation}
\tilde{\Pi}^{\tau}_{\alpha}=\Pi^{\tau}_{\alpha}
-\frac{i\pi}{2}\, K_{\alpha\beta}\,\tau_2^{-1}\cdot a^{\beta}_{\bar{\tau}}
\quad\text{and}\quad
\tilde{\Pi}^{\bar{\tau}}_{\alpha}=\Pi^{\bar{\tau}}_{\alpha}
+\frac{i\pi}{2}\, K_{\alpha\beta}\,\tau_2^{-1}\cdot a^{\beta}_{\tau}.
\end{equation}
We have the following commutation relations:
\begin{align}
&[\tilde{\Pi}^{\tau}_{p,\alpha},\,\tilde{\Pi}^{\bar{\tau}}_{q,\beta}]=
\pi\,(\tau_2^{-1})_{pq}K_{\alpha\beta};
\label{CCrel0}
\\
&[\mathscr{H}_{\text{harm}},\tilde{\Pi}^{\tau}_{p,\alpha}]=
-2\pi e^2\,\tilde{\Pi}^{\tau}_{p,\beta}(\lambda K)^{\beta}{}_{\alpha}
\quad\text{and}\quad
[\mathscr{H}_{\text{harm}},\tilde{\Pi}^{\bar{\tau}}_{p,\alpha}]=
2\pi e^2\,\tilde{\Pi}^{\bar{\tau}}_{p,\beta}(\lambda K)^{\beta}{}_{\alpha}
.
\label{CCrel1}
\end{align}
Now recall that to take advantage of the  K\"{a}hler structure on the configuration
space $\mathcal{C}$ we have to use the decomposition $V=V_+\oplus V_-$.
Notice that the matrix $K$ is positive definite on $V_+$ and negative
definite on $V_-$. Thus from \eqref{CCrel0} one concludes
that $\tilde{\Pi}^{\tau}_-$ and  $\tilde{\Pi}^{\bar{\tau}}_+$
are  creation operators while
$\tilde{\Pi}^{\tau}_+$ and $\tilde{\Pi}^{\bar{\tau}}_-$ are
annihilation operators. Equation \eqref{CCrel1} shows that the creation operators
increase energy by $2\pi e^2$ and annihilation decrease by the same amount.

Clearly the ground state function $\Psi$ must be annihilated by the annihilation
operators $\tilde{\Pi}^{\tau}_+$ and $\tilde{\Pi}^{\bar{\tau}}_-$.
This yields the following expression for $\Psi$:
\begin{equation}
\Psi(a_{\tau},a_{\bar{\tau}})=\exp\Bigl[-\frac{\pi}{2} g_K(a,a)
\Bigr]
\phi(a_{\tau}^-,\,a_{\bar{\tau}}^+).
\label{Grstate}
\end{equation}
Here $g_K$ is the metric \eqref{g_K} which is canonically
associated to the complex structure $\mathcal{J}$,
and $\phi(a_{\tau}^-,a_{\bar{\tau}}^+)$ is an arbitrary holomorphic
function of its variables. The energy of this ground state is
$E_{0}=\pi g e^2 \Tr(\lambda K \Gamma)/8$.
Comparing the expression above with the decomposition \eqref{holsplit}
one concludes that $\phi=\phi(a^{0,1})$ is an antiholomorphic function
on the configuration space $\mathcal{C}$. In section~\ref{sec:gauss} we have shown
that the physical wave function must satisfy the Gauss law constraint \eqref{Gausslaw}.
This constraint together with the holomorphy condition is very restrictive
and allows one to determine $\phi$ completely.

\subsection{The ground state as an antiholomorphic function}
\label{subsec:53}
In this section we write a very explicit basis of wavefunctions.
The final answer is given in equation \eqref{psiBeta} below.

\paragraph{Functional equation on $\phi$.}
Substituting \eqref{Grstate} into the Gauss law constraint
\eqref{Gausslaw} one finds that $\phi$ must satisfy
\begin{equation}
\phi(a^{0,1}+\omega^{0,1})=f^{\bullet}_{\sigma}(\omega)\,
e^{2\pi i\sum_r \int_{\Gamma_r}n^r(\omega)}
\,e^{\frac{\pi}{2}g_K(\omega,\omega)+
\pi g_K(\omega,a)-i\pi \omega_K(\omega,a)}\,\phi(a^{0,1})
\label{phi-law}
\end{equation}
where $\omega$ is a harmonic  element of the real space $\mathcal{C}_{\Lambda}$.
The function $\phi$ depends only on $a^{0,1}$ thus for this
expression to be well defined the exponential
must also depend only on $a^{0,1}$. Indeed, the combination
$g_K(\omega,a)-i\omega_K(\omega,a)=-2i\omega_K(\omega,a^{0,1})$.
Introduce the hermitian form $H$ on the complex vector space $\mathcal{C}^{0,1}$
and a factor of automorphy $a_{(H,\chi)}(\omega,a)$:
\begin{subequations}
\begin{equation}
H(u,v):=g_K(\overline{u^{0,1}},v^{0,1})-i\omega_K(\overline{u^{0,1}},v^{0,1})\quad
\text{and}\quad
a_{(H,\chi)}(\omega,a)=
\chi_{\sigma}(\omega)\,e^{\frac{\pi}{2}H(\omega,\omega)+
\pi H(\omega,a)}
\label{automorphy}
\end{equation}
where $\chi(\omega)$ is a map from $\mathcal{C}_{\Lambda}$ to $U(1)$:
\begin{equation}
\chi_{\sigma}(\omega)=f_{\sigma}^{\bullet}(\omega)\,
e^{2\pi i\sum_r \int_{\Gamma_r}n^r(\omega)}.
\end{equation}
\end{subequations}
In these terms equation
\eqref{phi-law} acquires the form
\begin{equation}
\phi(a^{0,1}+\omega^{0,1})=a_{(H,\chi_{\sigma})}(\omega^{0,1},a^{0,1})\,\phi(a^{0,1}).
\label{func.eq}
\end{equation}
This is a very well known equation \cite{CAV}. It determines global
antiholomorphic
sections of the line bundle $L_{(H,\chi_{\sigma})}$ over the \textit{complex}
torus $\mathcal{C}^{0,1}/\mathcal{C}^{0,1}_{\Lambda}$. The first Chern class
of the bundle $L_{(H,\chi)}$ is $[-\omega_K]$ and the cocycle $\chi_{\sigma}$
determines the holonomies. The holomorphic sections of this line bundle
can be expressed in terms of the theta functions.
A simple argument shows that there are $|\det K|^g$ different sections.
A more pedestrian method to obtain the wavefunctions involves writing an overcomplete set of functions
\eqref{Grstate} and projecting on the gauge invariant ones.
In this way one obtains a theta-series which admits a holomorphic
factorization (see appendix~\ref{app:GS}).

\paragraph{Explicit expression for the cocycle.}
After choosing a basis in $\mathcal{C}$, $\omega=m^1_p\alpha^p+m^p_2 \beta_p$,
we can write the cocycle $\chi_{\sigma}$ explicitly (see \eqref{fbullet:coord})
\begin{subequations}
\begin{equation}
\chi_{\sigma_{\mathscr{A}}+\epsilon}(\omega)=
\exp\Bigl\{
i\pi (m^1_p,m_2^p)
+2\pi i\bigl[\epsilon^1_p\otimes W_2+c^1_p\bigr](m_2^p)
-2\pi i \bigl[\epsilon_2^p\otimes W_2+c_2^p\bigr](m^1_p)
\Bigr\}
\label{Chi}
\end{equation}
where $\sigma_{\mathscr{A}}$ is the preferred spin structure
associated to the homology basis $\mathscr{A}$ (see explanation below).
The dependence on $\epsilon$ is obtained from the variational formula
\eqref{theSpin}. The terms $c^1_p$ and $c_2^p$ are elements of $V^*$ defined
by
\begin{equation}
c^1_p(\cdot)=\sum_{r=1}^{N_V}n^r(\cdot)\int_{\Gamma_r}\beta_p
\quad\text{and}\quad
c_2^p(\cdot)=-\sum_{r=1}^{N_V}n^r(\cdot)\int_{\Gamma_r}\alpha^p.
\label{c1c2}
\end{equation}
\label{cocycle1}
Notice that the cocycle \eqref{cocycle1}
depends only on the
homotopy classes of contours $[\Gamma_r]$ and classes $[\epsilon^1],[\epsilon_2]\in(\frac12\Zh/\Zh)^g$,
and $[W_2]\in\Lambda^*/2\Lambda^*$ but not on any particular
representative.
It is crucial to notice that the expression for the cocycle depends
on the choice of the marking $\mathscr{A}$
of the Riemann surface.
\end{subequations}

Let us recall some useful facts about spin structures on $\Sigma_g$.
The component
$\mathrm{Pic}_{g-1}$ of the Picard group consists of the
bundles of degree $g-1$. The Riemann vanishing theorem
can be used to characterize the zeros of theta function.
The $\Theta$ divisor is an analytic subset of the Jacobian
defined by equation $\theta(z,\tau)=0$.
It implies that for any homology basis $\mathscr{A}$ there is a preferred
spin bundle $S_{\mathscr{A}}\in \mathrm{Pic}_{g-1}$. If $P_0$ is a point on $\Sigma_g$ then
$I_{\mathscr{A}}[S_{\mathscr{A}}\otimes \mathscr{O}((1-g)P_0)]$ is point in the Jacobian
called the vector of Riemann constants \cite{mumford}.
Here $I_{\mathscr{A}}:\text{divisors}\to \mathrm{Jac}_{\mathscr{A}}(\Sigma_g)$ is the Abel map.

Let $\epsilon=(\epsilon^1,\epsilon_2)\in\frac12 \Zh^{2g}/\Zh^{2g}$ be the characteristics
of the theta function, then \cite{mumford,Atiyah}
\begin{equation}
(-1)^{\dim H^0(S_{\mathscr{A}}\otimes L_{\epsilon})}=(-1)^{4\epsilon^1\cdot\epsilon_2}.
\label{mod2index}
\end{equation}
Here $\dim H^0(S_{\mathscr{A}}\otimes L_{\epsilon})$  is the number of holomorphic
sections of the spin bundle $S_{\mathscr{A}}\otimes L_{\epsilon}$
($L_{\epsilon}$
is a line bundle with the flat connection determined by $\epsilon$).
Thus the left hand side is the  mod 2 index \cite{Atiyah}. The spin structure is called
even or odd depending on whether the mod 2 index is $+1$ or $-1$.

\paragraph{A basis in $\mathcal{C}^{0,1}$. The theta function.}
We choose a complex basis in $\mathcal{C}$:
\begin{equation}
C^{0,1}=\bigl(\alpha\otimes \check{e} K-\beta\otimes \check{e}\,\overline{
T}\,\bigr)(2i \Im T)^{-1}
\quad
\text{and}\quad C^{1,0}=\overline{C^{0,1}}
\label{c.basis}
\end{equation}
where $\check{e}^{\alpha}$ is the basis in $\Lambda^*$.
The complex period matrix $T$ is determined by
the equation
\begin{equation*}
\mathcal{J}C^{0,1}=-iC^{0,1}.
\end{equation*}
Explicitly:
\begin{equation}
T=-\bar{\tau}\otimes K P_+ -\tau\otimes KP_-
=-\tau_1\otimes K+i\tau_2\otimes \mu.
\label{periodT}
\end{equation}
An arbitrary harmonic element $\lambda=(\alpha\otimes \check{e} K)\cdot\lambda^1
+(\beta\otimes \check{e}K)
\cdot \lambda_2$ of the real vector space $\mathcal{C}$ can be written in the
complex basis \eqref{c.basis} as
\begin{equation}
\lambda=C^{0,1}\cdot\bigl[(\mathbbmss{1}_g\otimes K)\lambda_2+T\cdot\lambda^1\bigr]
+C^{1,0}\cdot\bigl[(\mathbbmss{1}_g\otimes K)\lambda_2+\overline{T}\cdot \lambda^1\bigr].
\label{c.decomp}
\end{equation}
Further we will denote $\mathbbmss{1}_g\otimes K$ by $\tilde{K}$.
This expression shows that the complex vector space $\mathcal{C}^{0,1}$ is
isomorphic to $T\Rh^{Ng}\oplus \tilde{K}\Rh^{Ng}$.
The lattice $\mathcal{C}_\Lambda$ is embedded into $\mathcal{C}^{0,1}$
according to this decomposition, and is isomorphic to
$T\Zh^{Ng}\oplus \tilde{K}\Zh^{Ng}$.
Substituting the
decomposition \eqref{c.decomp} into \eqref{automorphy} one finds
\begin{equation*}
H(\lambda,\omega)=(\lambda_2\cdot \tilde{K}+\lambda^1\cdot\bar{T})(\Im T)^{-1}
(\tilde{K}\cdot \omega_2+T\cdot\omega^1).
\end{equation*}
These data is already enough to solve the functional equation \eqref{func.eq} using
the Fourier transform.
The details can be found in \cite{CAV}, here we only present the result. Define
\begin{equation}
\vartheta_{\gamma}^{T c^1+\tilde{K} c_2}(T,v):=e^{\frac{\pi}{2}\,v\cdot(\Im T)^{-1}\cdot v
-i\pi (c^1_p,c_2^p)}
\sum_{\ell\in\Zh^{Ng}+\gamma+c^1}
e^{i\pi \ell\cdot T\cdot \ell +2\pi i (v+c_2K)\cdot\ell}.
\label{vartheta}
\end{equation}
Here $(c^1_p,c_2^p)=K_{\alpha\beta}\,c^{1,\alpha}_p
c^{p,\beta}_2$, $v=Tv_2+\tilde{K} v^1$,
$\gamma \in \tilde{K}^{-1}\Zh^{Ng}/\Zh^{Ng}$ is a representative
of the dual lattice,
and $c^1,c_2\in \Rh^{Ng}$ are characteristics of the theta function.
This canonical theta function satisfies the functional equation:
\begin{subequations}
\begin{equation}
\vartheta_{\gamma}^{T c^1+\tilde{K} c_2}(T,v+T \lambda^1+\tilde{K}\lambda_2)=
\chi_{c^1,c_2}(\lambda)\,e^{\frac{\pi}{2}H(\lambda,\lambda)+\pi H(\lambda,v)}\,
\vartheta_{\gamma}^{T c^1+\tilde{K}c_2}(T,v)
\end{equation}
where
\begin{equation}
\chi_{c^1,c_2}(\lambda)=e^{i\pi (\lambda^1_p,\lambda_2^p)+2\pi i (c^1_p,\lambda_2^p)
-2\pi i(c_2^p,\lambda^1_p)}
\end{equation}
\label{vartheta_cocycle}
\end{subequations}
The equations above become almost obvious if
one rewrites the canonical theta function \eqref{vartheta}  in the form
\begin{equation}
\vartheta_{\gamma}^{T c^1+\tilde{K} c_2}(T,v):=e^{\frac{\pi}{2}\,H(v,v)
+i\pi(v^1_p,v_2^p)-i\pi (c^1_p,c_2^p)}
\sum_{\ell\in\Zh^{Ng}+\gamma+c^1}
e^{i\pi (\ell+v^1)\cdot T\cdot (\ell+v^1) +2\pi i (v_2^p+c_2^p,\ell_p)}.
\label{vartheta2}
\end{equation}

\paragraph{The physical wave function.}
For the period matrix \eqref{periodT} the canonical theta function \eqref{vartheta} can be written in
terms of the Siegel-Narain theta function:
\begin{equation}
\Theta_{\Lambda+\gamma}^{\theta,\phi}(-\bar{\tau},P_{\pm};\xi)=e^{\frac{\pi}{2}\,
\mu_{\alpha\beta}\xi^{\alpha}\cdot\tau_2^{-1}\cdot\xi^{\beta}
-i\pi (\phi^p, \theta_p)}\sum_{\lambda\in\Lambda^{\otimes g}+\gamma+\theta}
e^{-i\pi\bar{\tau}^{pq} (\lambda_p,\lambda_q)_+
-i\pi\tau^{pq} (\lambda_p,\lambda_q)_- +2\pi i (\xi^p+\phi^p,\lambda_p)}
\label{SN-theta}
\end{equation}
where $\xi=(\xi_2-\bar{\tau}\xi^1)_++(\xi_2-\tau \xi^1)_-$,
$(\cdot,\cdot)$ denotes the bilinear form on $V$,
$(\lambda,\lambda)_{\pm}:=(P_{\pm}\lambda,\,P_{\pm}\lambda)$,
$\gamma$ is a representative of the discriminant group
$\Lambda^*/\Lambda$. $v$ and $c^1,c_2$  in \eqref{vartheta}
are related to $\xi$ and $\theta,\phi$ by $v=\tilde{K}\,\xi$, $c^1=\theta$ and $c_2=\phi$.

Comparing \eqref{cocycle1} with \eqref{vartheta_cocycle}
we can identify $\theta_p=\epsilon^1_p\otimes W_2
+c^1_p$ and $\phi^p=\epsilon_2^p\otimes W_2+c_2^p$
and write
\begin{equation}
\Psi_{\mathscr{A},\gamma,\epsilon}^{\text{phys}}(a^1,a_2,\tau,W_2;c^1,c_2)=
\Nc_g(\tau)\,e^{-\frac{\pi}{2}g_K(a,a)}
\,\Theta_{\Lambda+\gamma}^{\epsilon^1\otimes W_2+c^1,\,\epsilon_2\otimes W_2
+c_2}\bigl(-\bar{\tau},\,P_{\pm};
a_{\tau}^-+a_{\bar{\tau}}^+\bigr).
\label{psiBeta}
\end{equation}
Here $\gamma\in(\Lambda^*/\Lambda)^{\otimes g}$,
and $\epsilon^1,\epsilon_2\in(\frac12\Zh)^{\otimes g}$, $W_2\in\Lambda^*$
are \textit{particular} representatives of the classes $[\epsilon^1],\,[\epsilon_2]$
and $[W_2]$, $c^1,c_2$ are defined by \eqref{c1c2}.
The subscript $\mathscr{A}$ in the notation for the wave function
shows that it depends on the choice of marking $\mathscr{A}$.
$\Nc_g(\tau)$ is some normalization ``constant'' which we are going to fix later.
The wave functions \eqref{psiBeta} form a basis in the Hilbert space
of the ground states $\Hc_{\epsilon^1,\epsilon_2}(W_2)$.

It is interesting to notice that the wave function in the harmonic sector \eqref{psiBeta}
depends on the coupling constants $\lambda^{\alpha\beta}$
only through the matrix $\Gamma$ which determines the complex structure on $\mathcal{C}$.

The dependence of the wave function on a choice of representatives
$\epsilon^1,\epsilon_2\in(\frac12\Zh)^{\otimes g}$ of the classes $[\epsilon^1],\,[\epsilon_2]$ is summarized by
the following formula
\begin{equation}
\Psi_{\gamma,\epsilon^1+n^1,\epsilon_2+n_2}^{\text{phys}}=
e^{i\pi (W_2,W_2)[n^1_pn_2^p+\epsilon^1_p n_2^p
-\epsilon_2^p n^1_p]
+i\pi (W_2,\,n_2^pc^1_p-n^1_p c_2^p)
+2\pi in_2^p(\gamma_p,W_2)}
\,
\Psi_{\gamma+n^1\otimes \overline{W}_2,\,\epsilon^1,\epsilon_2}^{\text{phys}}.
\label{e-dep}
\end{equation}
Here $\overline{W}_2$ is the projection of $W_2$ onto the discriminant
group $\Lambda^*/\Lambda$.

The dependence of the wave function on a choice of representative
$W_2$ of the characteristic class $[W_2]\in \Lambda^*/2\Lambda^*$ is
\begin{equation}
\Psi_{\gamma,\,\epsilon^1,\epsilon_2}^{\text{phys}}(W_2+2\Delta)
=e^{4\pi i\epsilon^1_p\epsilon_2^p(\Delta,\Delta)
+4\pi i \epsilon_2^p(\gamma_p,\Delta)+
2\pi i\,(\Delta,c^1_p\epsilon_2^p -c_2^p\epsilon^1_p)
}\,
\Psi_{\gamma+2\epsilon^1\otimes\overline{\Delta},\,\epsilon^1,\epsilon_2}^{\text{phys}}(W_2)
\label{W-dep}
\end{equation}
where $\Delta$ is an arbitrary element of $\Lambda^*$.

\subsection{Magnetic translation group}
The generators \eqref{W}
of the magnetic translation group $\mathcal{W}$ are of the form
\begin{equation}
\mathtt{W}_a(\phi)=e^{i\phi^1_p\Pi_1^p+i\pi(\phi^1_p,a_2^p)-2\pi i (\phi^1_p,c_2^p)}
\quad\text{and}\quad
\mathtt{W}_b(\phi)=e^{i\phi^p_2\Pi_p^2-i\pi(\phi^p_2,a^1_p)+2\pi i (\phi_2^p,c^1_p)}.
\label{repW}
\end{equation}
Thus the magnetic translation group $\mathcal{W}$ acts on the physical
wave functions as follows:
\begin{subequations}
\begin{align}
\mathtt{W}_a(\phi)\Psi_{\gamma,\epsilon}^{\text{phys}}(a^1,a_2;\tau)
&=\Psi_{\gamma+\phi^1,\epsilon}^{\text{phys}}(a^1,a_2;\tau);
\\
\mathtt{W}_b(\phi)\Psi_{\gamma,\epsilon}^{\text{phys}}(a^1,a_2;\tau)
&=e^{2\pi i (\phi_2^p,\gamma_p)}\,
\Psi_{\gamma,\epsilon}^{\text{phys}}(a^1,a_2;\tau);
\end{align}
\label{Wrep}
\end{subequations}
where $\phi=\phi_{p}^{1}\,\alpha^p
+\phi_2^{p}\,\beta_p$ and $\phi^1,
\phi_2\in(\Lambda^*/\Lambda)^{\otimes g}$.
One sees that the space of the physical wave functions $\Hc_{\epsilon^1\epsilon_2}(W_2)$
is a representation space for the group $\mathcal{W}$.

{}From the representation \eqref{repW} one concludes that
the magnetic translation group $\mathcal{W}$ represents translations
of the theta function by the elements from the discriminant group.
Put differently, after fixing the complex structure $\mathcal{J}$ we are left
with finitely many holomorphically inequivalent
choices of embedding of the lattice $\mathcal{C}_{\Lambda}$
into the complex vector space $\mathcal{C}^{0,1}$. These choices
are parameterized by the discriminant group $\Lambda^*/\Lambda$.
The magnetic translation group is a permutation group
on the space of these embeddings.

\subsection{Normalization}
Since the Hamiltonian and Hilbert space factorize into flat and massive
sectors we can consider the wavefunction restricted to the flat fields.
The inner product on the space of flat fields is defined by
\begin{equation}
\langle\Psi_{\gamma,\sigma},\Psi_{\gamma',\sigma}\rangle:=
\int_{Z^{1}(\Sigma_g,V)}\frac{\Ds_g A^{\alpha}}{\mathrm{vol}(\text{gauge group})}\,
\overline{\Psi_{\gamma,\sigma}(A)}\,\Psi_{\gamma',\sigma}(A),
\label{inner}
\end{equation}
where the integral runs  over all closed $1$-forms on $\Sigma_g$
with values in $V$.
The integral descends to one on the space of gauge inequivalent flat fields.
This space is  ``the Jacobian'' $J=H^1(\Sigma_g,V)/
H^1(\Sigma_g,\Lambda)$.
It is interesting to notice that if we substitute the expression \eqref{Grstate}
into \eqref{inner} one obtains the natural inner product on the sections of
the line bundle $L(H,\chi_{\sigma})$.

We use the Hodge decomposition to write
$A=a^h+df$  where $a^h$ is a harmonic $1$-form and
$f$ is a function with values in $V$.
The measure as usual
can be obtained from the norm:
\begin{multline*}
\|\delta A^{\alpha}\|^2_g=\lambda^{-1}_{\alpha\beta}\int_{\Sigma_g}\delta A^{\alpha} *\delta A^{\beta}
=\lambda^{-1}_{\alpha\beta}\int_{\Sigma_g}\delta a_h^{\alpha}*\,\delta a_h^{\beta}
+\lambda^{-1}_{\alpha\beta}\int_{\Sigma_g}\delta f^{\alpha} *\Delta_0\delta f^{\beta}
\\
\quad\Rightarrow\quad
\Ds_g A^{\alpha}=(\det{}'\Delta_0)^{N/2}\,\Ds_g a_h^{\alpha}\,\Ds_g f^{\alpha}
\end{multline*}
where we used $\det\lambda=1$, and $\det{}'\Delta_0$ is the determinant of the
Laplacian operator $\Delta_0$ on the space of functions,
and ${}'$ means that we excluded
the zero modes.
The integral over $\Ds_g f$
cancels the volume
of the small gauge transformations.
The volume of the large gauge transformations is cancelled by
restricting the integral over $\mathrm{Harm}^1(\Sigma_g,V)$
to integral over  $J_{\mathscr{A}}=\mathrm{Harm}^1(\Sigma_g,V)/\mathrm{Harm}^1_{\Lambda}(\Sigma_g)$.
So after gauge fixing one obtains the following
expression for the $L^2$-norm:
\begin{equation}
\langle\Psi_{\gamma,\sigma},\Psi_{\gamma',\sigma}\rangle=
(\det{}'\Delta_0)^{N/2}\int_{J_{\mathscr{A}}}
\Ds_g a_h\,
\overline{\Psi_{\gamma,\sigma}(a_h)}\Psi_{\gamma',\sigma}(a_h).
\label{norm}
\end{equation}

Now we substitute \eqref{psiBeta} into \eqref{norm}.
Notice that
$a_2$ appears linearly in the exponential (this follows from \eqref{vartheta2}
and \eqref{SN-theta}), and
the integral over $a_2$ yields two Kronecker symbols $\delta_{\gamma\gamma'}$
and $\delta_{\lambda,\lambda'}$ where $\lambda,\lambda'\in\Lambda^{\otimes g}$
are summation variables in the definition \eqref{SN-theta} of the Siegel-Narain  theta function:
\begin{equation*}
\langle\Psi_{\gamma,\epsilon}^{\text{phys}},\Psi_{\gamma',\epsilon}^{\text{phys}}\rangle=
\delta_{\gamma,\gamma'}\,|\Nc_g(\tau)|^2\,
(\det{}'\Delta_0)^{N/2}
\sum_{n^{\alpha}_p\in\Zh}
\int_{0}^1 d a^{1,\alpha}_p\,
e^{-\pi\tau_2^{pq}(n_p+\gamma_p+\theta^1_p+a^1_p)^{\alpha}
\mu_{\alpha\beta}(n_q+\gamma_q+\theta^1_q+a^1_q)^{\beta}
}.
\end{equation*}
The sum over $n^{\alpha}$ combines with the integral over $(0,1)$
to give an integral over $\Rh$. This Gaussian
integral is easy to calculate and one finds:
\begin{equation}
|\Nc_g(\tau)|^2=|\det K|^{g/2}\,
\left(\frac{\det{}'\Delta_0}{\det\tau_2}\right)^{-N/2}.
\label{N(tau)}
\end{equation}
Evidently, from equation \eqref{N(tau)} $\Nc_g(\tau)$
is some kind of square root of the right hand side
of \eqref{N(tau)}. We now observe that there is a
very natural squareroot provided we view $\Nc_g(\tau)$
as a section of a line bundle rather than a function.
Notice that
the factor $\frac{\det{}'\Delta_0}{\mathrm{Vol}(\Sigma_g)\,\det\tau_2}$ is
the Quillen norm of the section $\det \bar{\pd}$ of the determinant
line bundle $\texttt{DET}(\bar{\pd})$ over the space of complex
structures on $\Sigma_g$:
\begin{equation*}
\frac{\det{}'\Delta_0}{\mathrm{Vol}(\Sigma_g)\,\det\tau_2}=\|
\det\pd
\|^2_Q=\|
\det\bar{\pd}
\|^2_Q
\end{equation*}
where $\mathrm{Vol}(\Sigma_g)$ is the volume
of the surface $\Sigma_g$ in the metric \eqref{metric}. Thus we can rewrite
\eqref{N(tau)} as
\begin{equation*}
|\Nc_g(\tau)|^2=|\det K|^{g/2}\,
\mathrm{Vol}(\Sigma_g)^{-N/2}\,\|(\det\bar{\pd})^{-r_-/2}(\det\pd)^{-r_+/2}\|^2_Q.
\end{equation*}
Here $r_\pm $ are the dimensions of $V_\pm$. Thus $r_+ + r_- = N$ and
$r_+ - r_- = \sigma $.

To obtain a holomorphic splitting we will consider $\Nc_g(\tau)$
as \textit{a section of the determinant bundle}
\begin{equation*}
\texttt{DET}(\bar{\pd})^{-r_+/2}\otimes\texttt{DET}(\pd)^{-r_-/2}
\end{equation*}
over the moduli space of complex structures $\mathscr{M}_{g,0}$.
The rationale for this choice is that we expect the dual conformal
field theory to have $r_+$ left-moving and $r_-$ right-moving bosons.
The norm of the wave function is now defined as a product of
the Quillen norm for $\Nc_g(\tau)$ and the $L^2$-norm for the theta function.
The normalization condition for the wave function $\|\Psi_{\gamma,\epsilon}\|^2=1$ implies
\begin{equation}
\Nc_g(\tau)=|\det K|^{g/4}\,
\mathrm{Vol}(\Sigma_g)^{-N/4}\,(\det\bar{\pd})^{-r_-/2}(\det\pd)^{-r_+/2}.
\label{N(g)}
\end{equation}
We can identify this expression with the partition
function of noncompact left- and right- moving bosons.

We will interpret $\det\pd$ as an automorphic form on
Teichm\"uller space.
Interpreted as a function on Teichm\"{u}ller space,
$\Nc_g(\tau)$ has the following modular properties:
\begin{equation*}
\Nc_g(-\tau^{-1})=\det(-i\tau)^{-r_-/2}\det(i\bar{\tau})^{-r_+/2}\Nc_g(\tau)
\end{equation*}
and
\begin{equation*}
\Nc_g(\tau+B)=e^{2\pi i(r_+-r_-)\phi(B)/24}\Nc_g(\tau)
\end{equation*}
 where
$B$ is a symmetric $g\times g$ matrix and
$\phi(B)$ is an integer.
We are being sloppy at this point. The line bundle $\texttt{DET}(\bar{\pd})$
has curvature, and there is no canonical trivialization over
Teichm\"uller space. We expect (given the results in \cite{Alvarez-Gaume:1987vm})
that there is a canonical isomorphism
of $\texttt{DET}(\bar{\pd})^{-r_+/2}\otimes\texttt{DET}(\pd)^{-r_-/2}$
with the line in which $\Theta$ is valued, allowing us to give a canonical
definition of $\Psi_{\mathscr{A},\gamma,\epsilon}^{\text{phys}}$ as a
function of $\tau$.

For genus $1$ we can be rather explicit.
Any metric on the torus can be written as
\begin{equation*}
ds^2=e^{2\phi(\sigma)}\,\frac{1}{\tau_2}|d\sigma^1-\tau d\sigma^2|^2
\end{equation*}
where $0\leqslant \sigma^i\leqslant 1$ and  $\tau$ is a complex number with positive
imaginary part. The determinant of the $\bar{\pd}$ operator is
\begin{equation*}
\det{}'\bar{\pd}=e^{\frac{1}{24\pi}S_L(\phi)}\eta(\tau)^2
\end{equation*}
where $S_L(\phi)$ is the Liouville action and $\eta(\tau)=q^{1/24}\prod_{n=1}^{\infty}
(1-q^n)$ is the Dedekind $\eta$-function. The Dedekind $\eta$-function has the following
modular properties: $\eta(-1/\tau)=(-i\tau)^{1/2}\eta(\tau)$
and $\eta(\tau+1)=e^{2\pi i/24}\eta(\tau)$. Thus,
\begin{equation}
\Nc_1(\tau)=e^{\frac{N}{48\pi}S_L(\phi)}
\biggl[\int_0^1 d^2\sigma\, e^{\phi(\sigma)}\biggr]^{-N/4}
\,\frac{|\det K|^{1/4}}{\eta^{\,r_-}(\tau)
\bar{\eta}^{\,r_+}(\bar{\tau})}.
\end{equation}

\subsection{Modular properties}
\label{sec:modular}
The action of the Teichm\"{u}ller modular group
on the wavefunctions factors through the action of the symplectic
modular group, defined by the action of large diffeomorphisms
on $H^1(\Sigma_g,\Zh)$. This group is isomorphic to $Sp(2g,\Zh)$.
The group $Sp(2g,\Zh)$ consist of the matrices of the form
\begin{equation}
g=\begin{pmatrix}
A & B
\\
C & D
\end{pmatrix}
\quad\text{and}\quad
D^tA-B^tC=\mathbbmss{1}_g,\quad
D^tB=B^tD,\quad C^tA=A^tC.
\end{equation}
It is generated by
\begin{enumerate}
\item $\begin{pmatrix} A & 0
\\
0 & A^{-1,t}
\end{pmatrix}$, $A\in GL(g,\Zh)$ i.e. $\det A=\pm 1$;

\item $\begin{pmatrix} \mathbbmss{1}_g & B \\ 0 & \mathbbmss{1}_g\end{pmatrix}$ where
$B$ is any symmetric integral $g\times g$ matrix;

\item $S=\begin{pmatrix} 0 & -\mathbbmss{1}_g\\ \mathbbmss{1}_g & 0\end{pmatrix}$.
\end{enumerate}

The modular properties of the Siegel-Narain theta-function \eqref{SN-theta} are as follows:

\begin{subequations}
\noindent$\bullet$ A-transform:
\begin{equation}
\Theta_{\Lambda+\gamma}^{\theta,\phi}(-A\bar{\tau} A^t,P_{\pm};A\xi)
=\Theta_{\Lambda+A^t\gamma}^{A^t\theta,A^{-1}\phi}(-\bar{\tau},P_{\pm};\xi)
\end{equation}

\noindent $\bullet$ B-transform (generalization of the T-transform):
\begin{multline}
\Theta_{\Lambda+\gamma}^{\theta,\phi}(-\bar{\tau}-B,P_{\pm};\xi)
=e^{\frac{i\pi}{2}B^{pp}(\theta_p,W_2)}\,
e^{-i\pi B^{pp}(\gamma_p,\gamma_p-W_2)
-2\pi i\sum_{p<q}B^{pq}(\gamma_p,\gamma_q)}\,
\\
\times
\Theta_{\Lambda+\gamma}^{\theta,\phi-B\cdot\theta-\frac12 \mathrm{diag}(B)\otimes W_2}
(-\bar{\tau},P_{\pm};\xi)
\label{Theta:T}
\end{multline}
where $W_2$ is a representative of the characteristic class
$[W_2]\in \Lambda^*/2\Lambda^*$. The left hand side of \eqref{Theta:T}
clearly does not depend on a particular choice of $W_2\in[W_2]$, thus
the right hand side also does not depend on this choice.
This freedom in the choice of a representative will be
important later when we will consider B-transformation of the wave function.

\noindent $\bullet$ $S$-transform:
\begin{multline}
\Theta_{\Lambda+\gamma}^{\theta,\phi}\bigl(\bar{\tau}^{-1},P_{\pm};\,\tau^{-1}\xi_++\bar{\tau}^{-1}\xi_-\bigr)
\\
=\det(-i\tau)^{r_-/2}\det(i\bar{\tau})^{r_+/2}\,|\Lambda^*/\Lambda|^{-g/2}
\sum_{\gamma'\in(\Lambda^*/\Lambda)^{\otimes g}}e^{2\pi i\,(\gamma_p,\gamma'_p)}
\,\Theta_{\Lambda+\gamma'}^{-\phi,\theta}(-\bar{\tau},P_{\pm};\xi).
\end{multline}
\label{Theta-mod}
\end{subequations}

Before applying these modular transformations
to the wave function \eqref{psiBeta} we need to introduce some notation.
The functions \eqref{psiBeta} are labeled by elements of the
discriminant group $\mathscr{D}=\Lambda^*/\Lambda$.
We denote by $b:\mathscr{D}\times \mathscr{D}\to \mathbb{Q}/\Zh$ the bilinear form on $\mathscr{D}$
inherited from the bilinear form on $\Lambda^*$.
Some quantities in the expressions below are naturally written in terms
of the functional $q_{W_2}$:
\begin{equation}
q_{W_2}(\gamma)=\frac12 (\gamma,\gamma-W_2)+\frac18(W_2,W_2) \mod 1.
\label{qW_2}
\end{equation}
where $(\cdot,\cdot)$ is the bilinear form on $\Lambda^*$. The fact that $W_2$
is a characteristic vector guarantees that $q_{W_2}$ descends
to a functional on the discriminant group. Moreover the functional $q_{W_2}$
does not change under the shifts $W_2\to W_2+2\lambda$ where $\lambda$
is an arbitrary element of the lattice $\Lambda$. The constant term in
\eqref{qW_2} was chosen in such a way that under the shifts
of $W_2$ by $2\Delta$, $\Delta\in \Lambda^*$ the functional \eqref{qW_2}
satisfies the equality
\begin{equation}
q_{W_2+2\Delta}(\gamma)=q_{W_2}(\gamma-\overline{\Delta}).
\label{w2D}
\end{equation}
Here $\overline{\Delta}$ denotes the projection of $\Delta$ onto the discriminant group.
Actually $q_{W_2}$ is a quadratic refinement of the bilinear form $b$
(see equation \eqref{q.ref} for the definition).
It is important to understand that the quadratic refinement
contains some extra information about the lattice which is
not encoded in the bilinear form on $\mathscr{D}$.
As we show below, if $b$ is nondegenerate then
every quadratic refinement of $b$ defines a distinguished
element in $\mathscr{D}$. For the case at hand it is $\overline{W}_2$ ---
the projection of $W_2$ onto the discriminant group.

Using the modular properties of the theta function \eqref{Theta-mod}
and the normalization section \eqref{N(g)} one can derive the
modular transformations of the wave function \eqref{psiBeta}:

\begin{subequations}
\noindent$\bullet$ $A$-transform:
\begin{equation}
\Psi_{\gamma,\epsilon^1,\epsilon_2}(\,A^ta^1,A^{-1}a_2,A\tau A^t;c^1,c_2)
=\Psi_{A^t\gamma,A^t\epsilon^1,A^{-1}\epsilon_2}(a^1,a_2,\tau;A^tc^1,A^{-1}c_2).
\end{equation}

\noindent$\bullet$ $B$-transform:
\begin{multline}
\Psi_{\gamma,\,\epsilon^1,\epsilon_2}(a^1,a_2+B a^1,\tau+B;c^1,c_2)
=e^{2\pi i(r_+-r_-)\phi(B)/24}\,e^{4\pi i \epsilon^1_p B^{pp}q_{W_2}(0)
+\frac{i\pi}{2} B^{pp}(c^1_p,W_2)
-2\pi i B^{pp}[q_{W_2}(\gamma_p)-q_{W_2}(0)]}\,
\\
\times
e^{-2\pi i\sum_{p<q} B^{pq}\,b(\gamma_p,\gamma_q)}\,
\Psi_{\gamma,\,\epsilon^1,\epsilon_2-B\epsilon^1-\frac12\mathrm{diag}(B)}(a_1,a_2,\tau;
c^1,c_2-Bc^1)
\label{preT}
\end{multline}
where $\phi(B)$ is some integer number which depends on the entries
of $B^{pq}$.
This phase arises from the factor $(\det\bar{\pd})^{-r_+/2}(\det \pd)^{-r_-/2}$ in $\Nc_g(\tau)$,
see equation \eqref{N(g)}, and
follows simply from the net gravitational
anomaly $c_L-c_R=r_+-r_-$ of $r_+$ left-moving and $r_-$ right-moving noncompact bosons.

\noindent$\bullet$ $S$-transform:
\begin{equation}
\Psi_{\gamma,\,\epsilon^1,\epsilon_2}(a_2,-a^1,-\tau^{-1};c^1,c_2)
=|\Lambda^*/\Lambda|^{-g/2}\sum_{\gamma'\in (\Lambda^*/\Lambda)^{\otimes g}}
e^{2\pi i\, b(\gamma_p,\gamma'_p)
}\,
\Psi_{\gamma',\,-\epsilon_2,\epsilon^1}(a^1,a_2,\tau;-c_2,c^1).
\label{preS}
\end{equation}
\label{modular}
\end{subequations}

To obtain the actual
$T$ and $S$ matrices one has to use \eqref{e-dep} for some values of
$\epsilon_2$. Evidently  expression \eqref{e-dep}  can be
rewritten in terms of the quadratic refinement $q_{W_2}$, \eqref{qW_2}:
\begin{equation}
\Psi_{\gamma,\epsilon^1+n^1,\epsilon_2+n_2}^{\text{phys}}=
e^{
8\pi i q_{W_2}(0)[n^1_pn_2^p+\epsilon^1_p n_2^p
-\epsilon_2^p n^1_p]
+i\pi (W_2,\,n_2^pc^1_p-n^1_p c_2^p)
+2\pi i n_2^p[q_{W_2}(-\gamma_p)-q_{W_2}(\gamma_p)]
}
\,
\Psi_{\gamma+n^1\otimes \overline{W}_2,\,\epsilon^1,\epsilon_2}^{\text{phys}}.
\label{e-dep-inv}
\end{equation}
Notice that only the term with the vortices
depends on $W_2$ not through the quadratic refinement.
{}From now on we will not consider the vortices. That is, we put $c^1=0,\, c_2=0$.

The main result of this discussion is that
the representation of the modular group (without vortices) is completely encoded in
the discriminant group $\mathscr{D}$, the bilinear form $b$ on the discriminant group,
the quadratic refinement $q_{W_2}$ and the signature $\sigma=r_+-r_-\mod 24$
of the bilinear form on the lattice $\Lambda$
modulo $24$.

\vspace{8mm}
\noindent\textbf{Lemma: } Two representations of the modular group
in the Hilbert spaces $\Hc(W_2)$ and $\Hc(W_2+2\Delta)$ are isomorphic.
The isomorphism is completely determined by the discriminant group $\mathscr{D}$,
the bilinear form $b$ and the quadratic refinement $q_{W_2}$.

\vspace{5mm}
\noindent \textbf{Proof: }
Here we consider only the representation of the modular group
without the vortices, so $c^1=0,\,c_2=0$.
The Hilbert space $\Hc(W_2)$ admits the following decomposition:
\begin{equation*}
\Hc(W_2)=\bigoplus_{\epsilon^1,\epsilon_2\in(\frac12\Zh/\Zh)^g}
\Hc_{\epsilon^1 \epsilon_2}(W_2)
\end{equation*}
where $\Hc_{\epsilon^1 \epsilon_2}$ is the Hilbert space
of wavefunctions with fixed spin structure.
We denote an operator $\Oc$ acting from  $\Hc_{\epsilon^{\prime\,1}\epsilon_2'}(W_2)$
to $\Hc_{\epsilon^1 \epsilon_2}(W_2)$ by
$\Oc(W_2)_{\gamma}{}^{\gamma'}\!
\left[
\begin{smallmatrix}
2\epsilon^{\prime\,1} & 2\epsilon'_2
\\
2\epsilon^1 & 2\epsilon_2
\end{smallmatrix}
\right]$.

The wave functions \eqref{psiBeta} form a basis in $\Hc_{\epsilon^1\epsilon_2}(W_2)$.
{}From equation \eqref{W-dep} we learn that the bases in the
Hilbert spaces $\Hc_{\epsilon^1\epsilon_2}(W_2+2\Delta)$ and
$\Hc_{\epsilon_1\epsilon_2}(W_2)$ are related
by the operator:
\begin{equation}
U(\Delta,\epsilon^1,\epsilon_2)_{\gamma}{}^{\gamma'}
=\delta_{\gamma',\gamma+2\epsilon^1\otimes\overline{\Delta}}\,\,
e^{4\pi i\epsilon_2^p\,b(\gamma_p,\overline{\Delta})}
e^{4\pi i\epsilon^1\cdot\epsilon_2(\Delta,\Delta)}.
\label{U}
\end{equation}
Notice that all factors except for the last one depend only on the projection $\overline{\Delta}$
of $\Delta$ onto the discriminant group.
If we shift $\Delta$ by an arbitrary element $\lambda$
of the lattice $\Lambda$ then the corresponding operators $U$ are related by
\begin{equation*}
U(\Delta+\lambda,\epsilon^1,\epsilon_2)=
\bigl[e^{4\pi i \epsilon^1\cdot\epsilon_2}\bigr]^{(\lambda,\lambda)}
U(\Delta,\epsilon^1,\epsilon_2).
\end{equation*}
The expression in the square brackets is the mod 2 index, see \eqref{mod2index}.
It equals $+1$ or $-1$ depending on whether the spin structure $\sigma_{\mathscr{A}}+
\epsilon^1_p\alpha^p+\epsilon_2^p\beta_p$
is even or odd. Thus if the spin structure is even
the operator \eqref{U} depends only on the data appearing in the lemma.
However if the spin structure is odd we have an ambiguity in the multiplication
by $\pm 1$.

The operators acting in the Hilbert spaces $\Hc(W_2+2\Delta)$
and $\Hc(W_2)$ are related by
\begin{equation*}
\Oc(W_2+2\Delta)_{\gamma}{}^{\gamma'}\!
\left[
\begin{smallmatrix}
2\epsilon^{\prime\,1} & 2\epsilon'_2
\\
2\epsilon^1 & 2\epsilon_2
\end{smallmatrix}
\right]
=U(\Delta,\epsilon^{1},\epsilon_2)_{\gamma}{}^{\gamma_1}
\Oc(W_2)_{\gamma_1}{}^{\gamma_2}\!
\left[
\begin{smallmatrix}
2\epsilon^{\prime\,1} & 2\epsilon'_2
\\
2\epsilon^1 & 2\epsilon_2
\end{smallmatrix}
\right]
(U^{-1})(\Delta,\epsilon^{\prime\,1},\epsilon'_2)_{\gamma_2}{}^{\gamma'}.
\end{equation*}
Substituting \eqref{U} one finds
\begin{multline}
\Oc(W_2+2\Delta)_{\gamma}{}^{\gamma'}\!
\left[
\begin{smallmatrix}
2\epsilon^{\prime\,1} & 2\epsilon'_2
\\
2\epsilon^1 & 2\epsilon_2
\end{smallmatrix}
\right]
=
\Oc(W_2)_{\gamma+2\epsilon^1\otimes\bar{\Delta}}{}^{\gamma'+2\epsilon^{\prime\,1}\otimes\bar{\Delta}}\!
\left[
\begin{smallmatrix}
2\epsilon^{\prime\,1} & 2\epsilon'_2
\\
2\epsilon^1 & 2\epsilon_2
\end{smallmatrix}
\right]
\\
\times
e^{2\pi i\,b(\overline{\Delta},2\epsilon_2\cdot\gamma-2\epsilon_2'\cdot\gamma')}
\,e^{2\pi i (\Delta,\Delta)\,\frac12[4\epsilon^1\cdot\epsilon_2
-4\epsilon^{\prime\,1}\cdot\epsilon_2']}.
\label{OOrel}
\end{multline}
Again all one on the right hand side depend only on the projection
$\overline{\Delta}$ of $\Delta$ onto the discriminant group
\textit{with one exception}, namely the last term.
The expression above is true for any operator. However the
operators which form a representation of the modular group have
a special property --- they preserve the mod 2 index.
Thus if $
\Oc\!
\left[
\begin{smallmatrix}
2\epsilon^{\prime\,1} & 2\epsilon'_2
\\
2\epsilon^1 & 2\epsilon_2
\end{smallmatrix}
\right]$
 is a modular transformation then the integer
$4\epsilon^{\prime\,1}\cdot\epsilon_2'
-4\epsilon^1\cdot\epsilon_2$ is an \textit{even} integer.
Hence we can change $(\Delta,\Delta)$ in the last
term of \eqref{OOrel} to $b(\overline{\Delta},\overline{\Delta})$.
So we prove that two representations of the modular group in
$\Hc(W_2+2\Delta)$ and $\Hc(W_2)$
are isomorphic and the isomorphism \eqref{OOrel} can be completely
written in terms of $\mathscr{D}$, $b$ and $q_{W_2}$. $\Box$

\subsubsection{Genus one}
For genus one it is easy to write the modular
matrices explicitly (without vortices).
As in the previous section
we denote an operator $\Oc$ acting from  $\Hc_{\epsilon^{\prime\,1}\epsilon_2'}$
to $\Hc_{\epsilon^1 \epsilon_2}$ by
$\Oc\!
\left[
\begin{smallmatrix}
2\epsilon^{\prime\,1} & 2\epsilon'_2
\\
2\epsilon^1 & 2\epsilon_2
\end{smallmatrix}
\right]$.
The wave functions \eqref{psiBeta} form a basis in $\Hc_{\epsilon^1 \epsilon_2}(W_2)$.
To extract $T$ and $S$ matrices one has to use \eqref{e-dep-inv}
for some values of $\epsilon^1,\epsilon_2$.
The nonzero $T$-matrices are:
\begin{subequations}
\begin{align}
T_{\gamma}{}^{\gamma'}\!
\left[
\begin{smallmatrix}
0 & 1
\\
0 & 0
\end{smallmatrix}
\right]
&=T_{\gamma}{}^{\gamma'}\!
\left[
\begin{smallmatrix}
0 & 0
\\
0 & 1
\end{smallmatrix}
\right]=
e^{2\pi i\sigma/24-2\pi i\, [q_{W_2}(-\gamma)-q_{W_2}(0)]}\,\delta_{\gamma\gamma'};
\\
T_{\gamma}{}^{\gamma'}\!
\left[
\begin{smallmatrix}
1 & 0
\\
1 & 0
\end{smallmatrix}
\right]
&=
T_{\gamma}{}^{\gamma'}\!
\left[
\begin{smallmatrix}
1 & 1
\\
1 & 1
\end{smallmatrix}
\right]
=e^{2\pi i\sigma/24-2\pi i\, q_{W_2}(-\gamma)}\,\delta_{\gamma\gamma'};
\end{align}
\label{T}
\end{subequations}
where $\sigma=r_+-r_-\mod 24$ is the signature of $\Lambda$ modulo $24$.
The nonzero $S$-matrices are:
\begin{subequations}
\begin{align}
S_{\gamma}{}^{\gamma'}\!
\left[
\begin{smallmatrix}
0 & 0
\\
0 & 0
\end{smallmatrix}
\right]
&=
S_{\gamma}{}^{\gamma'}\!
\left[
\begin{smallmatrix}
0 & 1
\\
1 & 0
\end{smallmatrix}
\right]
=
|\Lambda^*/\Lambda|^{-1/2}e^{2\pi i\, b(\gamma,\gamma')};
\\
S_{\gamma}{}^{\gamma'}\!
\left[
\begin{smallmatrix}
1 & 0
\\
0 & 1
\end{smallmatrix}
\right]
&=|\Lambda^*/\Lambda|^{-1/2}e^{2\pi i\, b(\gamma,\gamma'+\overline{W}_2)};
\\
S_{\gamma}{}^{\gamma'}\!
\left[
\begin{smallmatrix}
1 & 1
\\
1 & 1
\end{smallmatrix}
\right]
&=
|\Lambda^*/\Lambda|^{-1/2}e^{2\pi i\, b(\gamma,\gamma'+\overline{W}_2)
+4\pi iq_{W_2}(0)}
\end{align}
\label{S}
\end{subequations}
{}\!\!One can verify that the $S$ and $T$ operators presented above
satisfy the equivalence relation \eqref{OOrel}.
In the process of verification it is enough to use only the properties
of the quadratic refinement \eqref{q.ref} and \eqref{equiv}
but not the explicit expression \eqref{qW_2}.

\subsubsection{Some checks}
Abstractly the modular group in genus one without vortices is generated by two transformations
$S$ and $T$ subject to the relations
\begin{equation}
S^4=1,\quad (TS)^3=(ST)^3=S^2.
\end{equation}
Let us check that \eqref{T} and \eqref{S} satisfy these properties.

Explicit calculation shows that
\begin{alignat*}{2}
(S^2)_{\gamma}{}^{\gamma'}\!
\left[
\begin{smallmatrix}
0 & 0
\\
0 & 0
\end{smallmatrix}
\right]
&=\delta_{\gamma+\gamma',0};
&\qquad\qquad
(S^2)_{\gamma}{}^{\gamma'}\!
\left[
\begin{smallmatrix}
1 & 0
\\
1 & 0
\end{smallmatrix}
\right]
&=\delta_{\gamma+\gamma',0}\,e^{2\pi i\,b(\gamma,\overline{W}_2)};
\\
(S^2)_{\gamma}{}^{\gamma'}\!
\left[
\begin{smallmatrix}
0 & 1
\\
0 & 1
\end{smallmatrix}
\right]
&=
\delta_{\gamma+\gamma'+\overline{W}_2,0};
&\qquad\qquad
(S^2)_{\gamma}{}^{\gamma'}\!
\left[
\begin{smallmatrix}
1 & 1
\\
1 & 1
\end{smallmatrix}
\right]
&=
\delta_{\gamma+\gamma'+\overline{W}_2,0}\,
e^{8\pi i q_{W_2}(0)+2\pi i\, b(\gamma,\overline{W}_2)}
\end{alignat*}
where $\overline{W}_2$ denotes the projection of $W_2$ onto the discriminants group.
It easy to verify that $S^4=1$ for the first three operators above.
However $(S^4)\!\left[
\begin{smallmatrix}
1 & 1
\\
1 & 1
\end{smallmatrix}
\right]
=1$ requires the following equality
\begin{equation}
8\,q_{W_2}(0)=b(\overline{W}_2,\overline{W}_2)\mod 1
\end{equation}
which is certainly true for the quadratic refinement \eqref{qW_2}.

Straightforward calculation shows that the equation $TSTST=S$ (for all spin structures)
is equivalent to the following equality:
\begin{equation*}
e^{\frac{i\pi}{4}(r_+-r_-)}
|\Lambda^*/\Lambda|^{-1/2}\sum_{\gamma_2\in\Lambda^*/\Lambda}
e^{2\pi i [(\gamma_2,\gamma_1+\gamma_3)+q_{W_2}(0)- q_{W_2}(\gamma_2)
]}=e^{2\pi i q_{W_2}(-\gamma_1-\gamma_3)}.
\end{equation*}
After summation over $\bar{\gamma}=\gamma_1+\gamma_3$ we arrive at the equality:
\begin{equation}
|\Lambda^*/\Lambda|^{-1/2}\sum_{\gamma\in \Lambda^*/\Lambda}
e^{2\pi i\, q_{W_2}(\gamma)}\stackrel{?}{=}e^{2\pi i(r_+-r_-)/8}.
\label{sumff}
\end{equation}
In virtue of the relation \eqref{w2D} this sum does not depend on a particular choice of
representative $W_2$ of the class $[W_2]\in\Lambda^*/2\Lambda^*$.
Moreover it is easy to see that the absolute value of the sum in the left hand side
is $1$.
Actually equation \eqref{sumff} is one of the versions of the celebrated
Gauss-Milgram sum formula \cite{Milgram}. The original Gauss-Milgram sum formula
was proven for an even lattice for $q(\gamma)=\frac12 (\gamma,\gamma)$ by Milgram
(for a one dimensional lattice the formula \eqref{sumff} was
proven by Gauss and is called the Gauss sum formula). However
the shortest and most elegant proof was given by J.~Milnor and D.~Husemoller \cite{Milnor} (Appendix~4).
The proof of the Gauss-Milgram formula for the quadratic
refinement \eqref{qW_2} can be found in \cite{HopkinsSinger}
on page~70, Proposition~5.41.

\section{Summary: the classification theorem}
\label{sec:sum}
\setcounter{equation}{0}

Given an integral lattice $\Lambda$ we construct the following set of invariants:
\begin{enumerate}
\item $\sigma=(r_+-r_-)\mod 24$. The signature of the
quadratic form on $\Lambda$ mod $24$.

\item The discriminant group $\mathscr{D}=\Lambda^*/\Lambda$ with
the bilinear form $b:\mathscr{D}\times \mathscr{D}\to \mathbb{Q}/\Zh$
inherited from the bilinear form on $\Lambda^*$.

\item The equivalence class of the quadratic refinement $[q(\gamma)]$
of the bilinear form $b$
on $\mathscr{D}$. The quadratic refinement is a map $q:\mathscr{D}\to \mathbb{Q}/\Zh$
such that for any two elements $\gamma_1$ and $\gamma_2$ of $\mathscr{D}$
\begin{equation}
q(\gamma_1+\gamma_2)-q(\gamma_1)-q(\gamma_2)+q(0)=b(\gamma_1,\gamma_2).
\label{sum:q.ref}
\end{equation}
We say that two quadratic refinements $q_1$ and $q_2$
are equivalent if there exists
$\Delta\in \mathscr{D}$  such that for all $\gamma\in \mathscr{D}$
\begin{equation}
q_1(\gamma)=q_2(\gamma-\Delta).
\end{equation}
\end{enumerate}
These invariants satisfy the Gauss-Milgram constraint \eqref{sumff}:
\begin{equation}
|\mathscr{D}|^{-1/2}\sum_{\gamma\in \mathscr{D}}
e^{2\pi i\, q(\gamma)}=e^{2\pi i \sigma/8}.
\label{sum:constr}
\end{equation}
Note that this constraint fixes the ambiguity of adding a constant to $q$.
Notice too that this constraint does not depend on a choice of representative $q(\gamma)$
of the equivalence class $[q(\gamma)]$.

\vspace{8mm}
\noindent\textbf{Theorem: } A representation of the modular group of
a spin Chern-Simons theory defined by a lattice $\Lambda$ is
completely encoded in the invariants 1,2,3 of the lattice $\Lambda$
subject to the constraint \eqref{sum:constr}.

\vspace{5mm}
\noindent\textbf{Proof: }
The proof of this statement was already indicated in section~\ref{sec:modular}.
Here we only have to show that every representative of the equivalence class
of quadratic refinements  $[q(\gamma)]$ is of the form $q_{W_2}$, see equation \eqref{qW_2}.
Once this is established, we can use equations \eqref{modular} and \eqref{e-dep-inv}
to extract the projective representation of the modular group.
All these expressions depend only on $\mathscr{D}$, $b$ and $q_{W_2}$.
{}From the lemma in section~\ref{sec:modular} it follows
that the representation of the modular group corresponding
to different choices of $W_2\in [W_2]$ are isomorphic.
Thus the theorem follows. $\Box$

Now let us explain why any quadratic refinement subject
to the constraint \eqref{sum:constr} is necessarily of the form $q_{W_2}$.
There is a theorem which states that the quadratic refinements $q_0$ with
$q_0(0)=0$ are in one-to-one correspondence with the projections $\overline{W}_2$
of representatives $W_2$ of the characteristic class $[W_2]$ onto the discriminant group.
The idea of the proof is as follows. Suppose we are given a quadratic
refinement $q_0(\gamma)$.
Then there exists a distinguished element of $\mathscr{D}$.
Indeed, consider the map $L:\mathscr{D}\to \mathbb{Q}/\Zh$
\begin{equation*}
L(\gamma)=q_0(-\gamma)-q_0(\gamma).
\end{equation*}
{}From \eqref{sum:q.ref} it follows that $L$ is a linear map. Since this
map is defined on $\mathscr{D}$ and $b$ is nondegenerate it follows that there exists an element $\overline{W_2}\in \mathscr{D}$
such that $L(\gamma)=b(\overline{W}_2,\gamma)$.

Now for the converse: Given a representative $W_2$
of the characteristic class $[W_2]$ we construct the quadratic refinement
\begin{equation}
q_0(\gamma)=\frac12(\gamma,\gamma-W_2)\mod 1.
\end{equation}
This functional is invariant under a shift of $W_2$ by $2\lambda$ where $\lambda\in\Lambda$.
Thus $q_0(\gamma)$ does not depend on a particular choice of the lift
$W_2\to\overline{W}_2\in \mathscr{D}$.
An arbitrary quadratic refinement is of the form $q_0(\gamma)+\ell(\gamma)+c$
where $\ell(\gamma)$ is a linear map and $c$ is a constant. Using that $b$
is nondegenerate on $\mathscr{D}$ we can represent $\ell(\gamma)=b(\gamma,\bar{\Delta})$
for $\bar{\Delta}\in \mathscr{D}$.
Now it can be shown that the addition of a linear function $\ell$ corresponds
to a shift $W_2\mapsto W_2-2\Delta$ where $\Delta$ is an arbitrary
lift of $\bar{\Delta}$ to $\Lambda^*$.
This shift is just the change of a representative of the class $[W_2]$.
Thus the only real ambiguity is the constant $c$.
It can be fixed from the Gauss-Milgram sum formula \cite{HopkinsSinger}:
\begin{equation}
|\Lambda^*/\Lambda|^{-1/2}\sum_{\gamma\in|\Lambda^*/\Lambda|}
e^{2\pi i q_0(\gamma)}=e^{2\pi i (\sigma-(W_2,W_2))/8}.
\end{equation}
We are looking for the quadratic
refinement which satisfies the constraint
\eqref{sum:constr}, thus $c=\frac18 (W_2,W_2)\mod 1$.
So as we expected  $q_0(\gamma)+c=q_{W_2}(\gamma)$.

\vspace{8mm}
\noindent\textbf{Corollary: } Two quantum spin Chern-Simons theories
defined by the lattices $\Lambda_1$ and $\Lambda_2$
are equivalent (have the same $3$-manifold invariants) iff
the invariants 1,2,3 for the lattices $\Lambda_1$ and $\Lambda_2$
are equal.

\vspace{5mm}
\noindent\textsc{Example 1. }
As simple examples, the three invariants vanish for the even unimodular lattice $II^{1,1}$.
This was the case studied in \cite{Witten:2003ya}. Another example is
 the odd unimodular lattice $I^{1,1}$ where again the invariants 1,2,3 vanish.
Thus the Chern-Simons theory determined by this lattice is trivial. Indeed,
the conformal field theory dual is a single left-moving and right-moving
Weyl fermion. The partition function on a Riemann surface with fixed spin
structure is clearly invariant under modular transformations preserving the spin structure.

\vspace{5mm}
\noindent\textsc{Example 2. }
The quantum Chern-Simons theories defined by
the lattices $\Lambda$ and $\Lambda\oplus (I^{1,1})^d$
are equivalent.

\vspace{5mm}
\noindent\textsc{Example 3. }
Consider the $E_8$ lattice.
The $E_8$ lattice is an even unimodular lattice and thus
its discriminant group is trivial, but it has signature $\sigma=8$
which is nonzero mod $24$. So this theory is nontrivial.

\vspace{5mm}
\noindent\textsc{Example 4. }
Any even unimodular lattice of signature $0\mod 24$ is trivial.

\vspace{5mm}
This result raises the natural question of whether there is a converse result.

\noindent\textbf{Theorem (converse result): }
Every quartet: an integer $\sigma$ modulo $24$,
a finite abelian group $\mathscr{D}$ together with a bilinear form $b:\mathscr{D}\times \mathscr{D}\to \mathbb{Q}/\Zh$,
and an equivalence class $[q]$ of a quadratic refinement of $b$
subject to the Gauss-Milgram constraint \eqref{sum:constr}
is realized by some lattice $\Lambda$.
Moreover if there exists a representative $q\in[q]$ such that $q(0)=0$
then the quartet is realized by an \textit{even} lattice.

\vspace{5mm}
\noindent\textbf{Corollary: }a) The set of quantum
\textit{spin} Chern-Simons theories is in one-to-one
correspondence with these quartets subject to the Gauss-Milgram constraint.
\\
b) The set of quantum \textit{integral} Chern-Simons
theories is in one-to-one correspondence with
the quartets subject to the Gauss-Milgram constraint
such that there exists a representative $q$ of the class $[q]$ with $q(0)=0$.

\vspace{5mm}
\noindent\textbf{Proof: } Suppose we are given the quartet
$(\sigma\mod 24,\,\mathscr{D},\,b,\,[q])$.
First, from Corollary~1.16.6 in \cite{Nikulin} it follows
that for sufficiently large $r_{\pm}\geqslant 0$ there
exists an odd lattice of signature $(r_+,r_-)$ which produces
the discriminant form $b$. Thus we can always
realize the triplet ($r_+-r_-=\sigma\mod 24,\,\mathscr{D},b$) by some
odd lattice. Now we can use the fact that
given a lattice the quadratic refinement $q$ is necessarily of the form
$q_{W_2}$. Thus the theorem follows.

Second, suppose that there exists $q\in[q]$ such that $q(0)=0$.
{}From Corollary~1.10.2 in \cite{Nikulin} it follows
that for sufficiently large $r_{\pm}$ subject
to the constraint $r_+-r_-=\mathrm{sgn} (q)\mod 8$
there exists an \textit{even} lattice of the signature $(r_+,r_-)$
which produces the quadratic refinement $q$.
One of the ways to define the signature of the quadratic
refinement (for an even lattice) (see Appendix~4 in \cite{Milnor})
 is to use the Gauss-Milgram sum formula
\eqref{sum:constr}. The $\sigma$ on the right hand side is the signature
of $q$ modulo 8. Thus in our case the condition of
the Corollary~1.10.2 is automatically
satisfied.
Therefore the
quartets for which there exists a representative $q\in [q]$
such that $q(0)=0$ are realized by \textit{even} lattices.
$\Box$

\vspace{5mm}
\noindent\textsc{Example 5. }
As an example of the converse result consider the
group $\mathscr{D}=\Zh/3\Zh$. There are two bilinear forms we can
define on $\mathscr{D}$: $b^{(1)}(x,y)=\frac{xy}{3}\mod 1$ and $b^{(2)}(x,y)=\frac{2xy}{3}\mod 1$.
One can easily solve equation \eqref{q.ref} for the quadratic refinement.
One finds that for each bilinear form there is only one equivalence
class of the quadratic refinements. The representatives which satisfy the Gauss-Milgram
constraint \eqref{constraint} are
\begin{equation}
q^{(1)}_{\sigma}(x)=q_0^{(1)}(x)+\frac{\sigma+2}{8}\mod 1\quad
\text{and}\quad
q^{(2)}_{\sigma}(x)=q_0^{(2)}(x)+\frac{\sigma-2}{8}\mod 1
\label{qexamp}
\end{equation}
where
\begin{center}
\renewcommand{\arraystretch}{1.2}
\begin{tabular}{c|ccc}
$x$ & $\bar{0}$ &$\bar{1}$ &$\bar{2}$ \\
\hline
$q_0^{(1)}$ &
$0$ & $\frac23$ & $\frac23$
\end{tabular}
\qquad\text{and}\qquad
\begin{tabular}{c|ccc}
$x$ & $\bar{0}$ &$\bar{1}$ &$\bar{2}$ \\
\hline
$q_0^{(2)}$ &
$0$ & $\frac13$ & $\frac13$
\end{tabular}
\end{center}
An odd lattice exists for any signature $\sigma\mod 8$. For example,
the quartet $(\sigma=1\mod 8,\,\mathscr{D},\, b^{(1)},\,[q^{(1)}_1])$
is realized by the lattice $\Zh\langle e_1\rangle$ with
$(e_1,e_1)=3$; the quartet
$(\sigma=0\mod 8,\,\mathscr{D},\, b^{(2)},\,[q^{(2)}_0])$
is realized by the lattice $\Zh\langle e_1,e_2\rangle$
with  $(e_i,e_j)=\left(\begin{smallmatrix}
1 & 2
\\
2 & 1
\end{smallmatrix}\right)$.

Even lattices usually do not exist for all signatures.
The quartet is realized by an even lattice if
there is a representative $q$ of $[q]$ such that $q(\bar{0})=0\mod1$.
Equations $q_{\sigma}^{(1)}(\bar{0})=0$ and
$q_{\sigma}^{(2)}(\bar{0})=0$ for the quadratic
refinement \eqref{qexamp} yield $\sigma=-2$ and $\sigma=2$ respectively
(compare with Table~15.4 in \cite{Conway}).
Thus the quartets  $(\sigma=-2\mod 8,\,\mathscr{D},\, b^{(1)},\,
[q^{(1)}_{-2}])$ and $(\sigma=2\mod 8,\,\mathscr{D},\, b^{(2)},\,
[q^{(1)}_{2}])$ are realized by even lattices.
For example, the second quartet is realized
by the lattice $\Zh\langle e_1,e_2\rangle$
with  $(e_i,e_j)=\left(\begin{smallmatrix}
2 & 1
\\
1 & 2
\end{smallmatrix}\right)$.

\paragraph{Monoid structure.}
The space of quantum Chern-Simons theories has
a structure of monoid. The sum operation is defined by
\begin{equation*}
(\sigma_1\mod 24,\,\mathscr{D}_1,\,b_1,\,[q_1])
\oplus (\sigma_2\mod 24,\,\mathscr{D}_2,\,b_2,\,[q_2])
=(\sigma_1+\sigma_2\mod 24,\,\mathscr{D}_1\oplus\mathscr{D}_2,\,b_{12},\,[q_{12}])
\end{equation*}
where the bilinear form $b_{12}$ and the quadratic refinement $q_{12}$
are defined by
\begin{align*}
b_{12}(x_1\oplus x_2,\,y_1\oplus y_2)&=b_1(x_1,x_2)+b_2(x_2,y_2)\mod 1
\\
q_{12}(x_1\oplus x_2)&=q_1(x_1)+q_2(x_2)\mod 1.
\end{align*}
Evidently the quadratic refinement defined in this way
satisfies the Gauss-Milgram constraint with the signature $\sigma_1+\sigma_2\mod 8$.
The identity element is the trivial quartet.

One can ask whether the space of quantum Chern-Simons
theories has in addition an abelian group structure. It is well know that
the space of quadratic refinements (with nondegenerate $b$) over a commutative ring
with identity form a Witt ring (see Appendix~1 in \cite{Milnor}).
The sum operation is defined in the same way as above,
while the subtraction operation is defined using the Grothendieck construction.
The trivial elements are so-called split modules.
For the case at hand the ring is $\Zh$. The quartet
$(\sigma,\,\mathscr{D},\,b,\,[q])$ is called split if there
exists a $\Zh$-submodule $N$ of $\mathscr{D}$ such that $N^{\perp}=N$ and $q(N)=q(0)+b(N,\Delta)$
for some $\Delta\in \mathscr{D}$. Here $N^{\perp}$ is a set of all $x\in \Ds$
such that $b(x,n)=0$ for all $n\in N$. Notice that in this definition
the nondegeneracy of $b$ is important.
Notice also that the definition of the split quartet
does not depend on a particular choice of quadratic refinement from $[q]$.
Two quartets $X_i=(\sigma_i,\mathscr{D}_i,b_i,[q_i])$  $i=1,2$
are called equivalent if there exist two split quartets
$S_i=(s_i,\mathcal{S}_i,\tilde{b}_i,[\tilde{q}_i])$
such that $X_1\oplus S_1\cong X_2\oplus S_2$.
The zero element is the equivalence class of split quartets.
However these split quartets correspond to \textit{non trivial} quantum
Chern-Simons theories (for example $|\mathcal{S}|>1$). For this reason we can not impose
this equivalence relation. Hence the quantum
Chern-Simons theories do not  seem to have an abelian group structure.

\section{Fractional Quantum Hall Effect}
\label{sec:QFHE}
\setcounter{equation}{0}
This section is based on Refs. \cite{Read,Frohlich:1991wb,Wen,Zee}.
A quantum Hall fluid in the long-distance limit has an effective
description in terms of \textit{spin} Chern-Simons theory with Maxwell kinetic term.
This is precisely the theory we consider in this paper.
The field strengths $F^{\alpha}$ play the role of electric currents
associated with
the $\alpha$-th Landau level. In the long distance limit these
currents interact very weakly, namely,
only through the Chern-Simons term. Let $\mathcal{A}$ be the gauge potential
of the external electro-magnetic field. We assume that it is
topologically trivial. The currents couple to the external field in the
standard way: $\int_{X_3}\mathcal{A}\wedge F^{\alpha}Q_{\alpha}$ where $Q_{\alpha}$ is
a set integers. The integers $Q_{\alpha}$ must be very special in order to
describe couplings to the Landau levels (see below).
We are interested in the following partition function
\begin{equation}
\mathcal{Z}(\mathcal{A})=\int\frac{\Ds A^{\alpha}}{\mathrm{Vol}(\text{gauge})}\,
e^{2\pi i \CS_{X_3}^{\text{spin}}(A^{\alpha})}
\mathrm{Hol}_{A^{\alpha}}(\{C_r,n^r\})\,
e^{2\pi i\int_{X_3} \mathcal{A}\wedge F^{\alpha}Q_{\alpha}}
e^{\text{kinetic}}
\label{Z[B]}.
\end{equation}

The variation with respect to $\mathcal{A}$ yields the expectation value
of the electro-magnetic current $J_{\text{em}}$ in the system:
\begin{equation}
\frac{1}{2\pi i}\frac{\delta}{\delta \mathcal{A}}\log \mathcal{Z}(\mathcal{A})=
\langle J_{\text{em}}\rangle\quad\text{where}\quad
J_{\text{em}}=Q_{\alpha}F^{\alpha}.
\end{equation}
This expectation value is easy to calculate. Since $\Ac$ is
topologically trivial one can shift the integration
variable $A^{\alpha}\mapsto A^{\alpha}-(K^{-1})^{\alpha\beta}Q_{\beta}\,\Ac$:
\begin{equation*}
\mathcal{Z}(\Ac)=e^{i\pi(Q,Q)\int_{X_3}\Ac\wedge d\Ac+2\pi i
\sum_{r=1}^{N_V}(n^r,Q)\int_{C_r}\Ac}\,
\mathcal{Z}(0)
\end{equation*}
where $Q=Q_{\alpha}\check{e}^{\alpha}\in\Lambda^*$.
Thus the expectation value of the electric current is
\begin{equation}
\langle J_{\text{em}}\rangle=(Q,Q)\,F_{\Ac}
+\sum_{r=1}^{N_V}(n^r,Q)\,\delta^{(2)}(x-x_r).
\end{equation}
The Hall conductivity is $\sigma_H=(Q,Q)$.
The total electric charge is given by the integral of this current
over the surface. The quantity $(Q,n^r)$ is naturally associated to the electric charge
of the vortex $n^r$. Notice that both $Q$ and $n^r$ are elements of
the dual lattice $\Lambda^*$, thus generically the electric charge is fractional.
The statistics of the vortex $n^r$ is described by the product $(n^r,n^r)$
\cite{Zee}.
Hence in general the vortices $n^r$ have fractional statistics. However if $n^r$
is an element of the lattice then both its electric charge and
statistics are integral. Certainly we want
the excitations with an odd charge  (such as electrons)
to have Fermi statistics, and the excitations with an even charge
to have Bose statistics. In other words we must require that
\begin{equation}
\text{for all } n\in\Lambda\quad (n,n)=(Q,n)\mod 2.
\end{equation}
Thus $Q$ is actually a representative of the
characteristic class $[W_2]\in \Lambda^*/2\Lambda^*$.

One of the invariants of spin Chern-Simons theory is
the quadratic refinement $q(\gamma)$ of the bilinear form on the discriminant group
$\Lambda^*/\Lambda$. For a particular choice of basis it reads
\begin{equation}
q_{W_2}(\gamma)=\frac12\,(\gamma,\gamma-W_2)+\frac18\, (W_2,W_2)\mod 1.
\end{equation}
One immediately observes that $\sigma_H=8q_{Q}(0)\mod 8$.
Now recall that two quadratic refinements $q$ and $q'$ are equivalent if
there exists $\Delta\in \Lambda^*/\Lambda$ such that $q'(\gamma)=q(\gamma-\Delta)$.
For the case at hand this means $q_{W_2+2\Delta}(\gamma)=q_{W_2}(\gamma-\Delta)$.
The set of all possible fractional Hall conductivities
associated with a given lattice $\Lambda$ is thus constructed  by:
\begin{equation}
\{\sigma_H\mod 8\}=\{(Q,Q)\mod 8\,|\,Q\in[W_2]\}
=\{8q(\gamma)\mod 8\,|\,\gamma\in\Lambda^*/\Lambda\}.
\end{equation}

\begin{figure}[t]
\centering
\includegraphics[width=230pt]{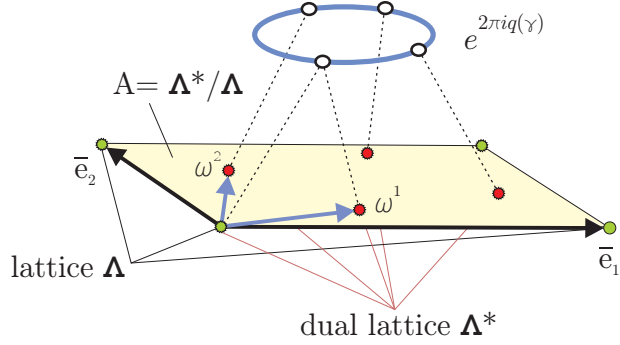}
\caption{The fundamental domain of the dual lattice $\Lambda^*$
for the Example~5. The basis vectors of the dual lattice are $\omega^1$ and $\omega^2$.
The lattice $\Lambda$ is spanned by the vectors $\bar{e}_1,\bar{e}_2$.
The red dots together with the origin represent the discriminant group $A$.
The circle on the top represents the values of the quadratic refinement
$q_{\omega^1}$.}
\label{fig:lattice}
\end{figure}

\vspace{5mm}
\noindent\textsc{Example 5. }
Consider a two dimensional lattice
$\Lambda=\Zh\langle e_1,e_2\rangle$ with the bilinear form
$(e_{\alpha},e_{\beta})=\left(\begin{smallmatrix}
3 & -1
\\
-1 & 2
\end{smallmatrix}
\right)$ (see Figure~\ref{fig:lattice}). The dual lattice $\Lambda^*$
is generated by vectors $\omega^1,\omega^2$ which
are defined by $\omega^{\beta}(e_{\alpha})=\delta^{\beta}{}_{\alpha}$.
We identify the lattice $\Lambda\subset\Lambda^*$
as $\Lambda\cong\Zh\langle\bar{e}_1,\bar{e}_2\rangle$ where $\bar{e}_{\alpha}=(e_{\alpha},\cdot)$.
The discriminant group is a cyclic group of dimension $5$.
It is represented by the classes
\begin{equation*}
\gamma_0=[0],\quad \gamma_1=[\omega^1],\quad \gamma_2=[2\omega^1],\quad
\gamma_3=[3\omega^1],\quad \gamma_4=[4\omega^1].
\end{equation*}
The characteristic vector $W_2$ can be chosen to be $\omega^1$. The values of
the quadratic refinement and the corresponding fractional Hall conductivities are
\begin{equation*}
\{q_{\omega^1}(\gamma_i)\}=\left\{\tfrac{1}{20},\tfrac{1}{20},
\tfrac{9}{20},\tfrac{5}{20},\tfrac{9}{20}
\right\}\!\!\mod 1
\quad\text{and}\quad
\Sigma_H=\{8\,q(\gamma_i)\!\mod 8\}
=\left\{\tfrac{2}{5},\tfrac{2}{5},3\tfrac{3}{5},2,3\tfrac{3}{5}\right\}
\!\!\mod 8.
\end{equation*}
The fractional electric charge $Q_{el}(\gamma)=q(-\gamma)-q(\gamma)\mod 1$:
\begin{equation*}
\{Q_{el}(\gamma_i)\}=\{0,\tfrac25,\tfrac45,\tfrac15,\tfrac35\}.
\end{equation*}

\section*{Acknowledgments}
We would like to thank M.~Douglas, D.~Freed, L.~Ioffe, J.~Jenquin, P.~Landweber,
S.~Lukic,  V.~Nair,
N.~Read, R.~Stong,   L.~Takhtadjan, and E.~Witten for discussions and correspondence.
This work was supported  in part by DOE grant DE-FG02-96ER40949,
D.B. was supported in part by RFBR grant 02-01-00695.

\clearpage
\appendix
\renewcommand {\theequation}{\thesection.\arabic{equation}}

\section{Gaussian sums on torus}
\label{app:GS}
\setcounter{equation}{0}
Here we want to solve the following functional equation
\begin{equation}
\phi(a^{0,1})=\chi_{c^1,c_2}^*(\omega)\,
\,e^{-\frac{\pi}{2}g_K(\omega,\omega)
-\pi g_K(\omega,a)+i\pi \omega_K(\omega,a)}
\phi(a^{0,1}+\omega^{0,1})
\label{B:phi}
\end{equation}
where the cocycle $\chi_{c^1,c_2}(\omega)$ is ($\omega=m^1_p\alpha^p+m^p_2 \beta_p$)
\begin{equation}
\chi_{c^1,c_2}(\omega)=
\exp\Bigl\{
i\pi (m^1_p,m_2^p)
+2\pi i(c^1_p,m_2^p)
-2\pi i (c_2^p,m^1_p)
\Bigr\}.
\label{B:chi}
\end{equation}
First we choose an over complete basis of antiholomorphic functions
\begin{equation*}
\phi(v^{1,0},a^{0,1}):=e^{\pi g_K(v,a)-i\pi\omega_K(v,a)}.
\end{equation*}
Second, using \eqref{B:phi} we average this function over the large gauge transformation:
\begin{align}
S(v^{1,0},a^{0,1})&=\sum_{\omega\in\mathcal{C}_{\Lambda}}\chi^*(\omega)\,
\,e^{-\frac{\pi}{2}g_K(\omega,\omega)
-\pi g_K(\omega,a)+i\pi \omega_K(\omega,a)}
\phi(v,a)
\notag
\\
&=e^{-2\pi i\omega_K(v^{1,0},a^{0,1})}\sum_{\omega\in\mathcal{C}_{\Lambda}}
\chi^*(\omega)e^{-\frac{\pi}{2} g_K(\omega,\omega)+2\pi i \omega_K(\omega,a^{0,1}+v^{1,0})}.
\label{B:phi2}
\end{align}
We now follow
a standard procedure and use Poisson resummation to split this sum
in a form so that we can
extract the conformal blocks. The details are presented below.
Using \eqref{B:result} one finds that the sum \eqref{B:phi2} can be written as
\begin{equation}
S(v^{1,0},a^{0,1})=
\frac{(\det 2\tau_2)^{N/2}}{|\det K|^{g/2}}\,
\sum_{\gamma\in(\Lambda^*/\Lambda)^{\otimes g}}
\Theta^{c^1,-c_2}_{\Lambda+\gamma}\bigl(\tau,P_{\pm};\,-v^{1,0}\bigr)
\Theta^{c^1,c_2}_{\Lambda+\gamma}\bigl(-\bar{\tau},P_{\pm};\,a^{0,1}\bigr)
\end{equation}
Thus identifying the cocycle \eqref{B:chi} with \eqref{Chi}
one finds that the physical wave function is
given by equation \eqref{psiBeta}.

\paragraph{Splitting the sum.}
We define the Siegel-Narain theta function
\begin{multline}
\Theta^{\theta,\phi}_{\Lambda+\gamma}(\tau,P_{\pm};\xi)=e^{
\frac{\pi}{2}\,
\mu_{\alpha\beta}\xi^{\alpha}\cdot\tau_2^{-1}\cdot\xi^{\beta}-i\pi(\phi^p,\theta_p)}\,
\\
\times
\sum_{\lambda\in\Lambda^{\otimes g}+\gamma+\theta}
\exp\Bigl[
i\pi\tau^{pq}(\lambda_p,\lambda_q)_+
+i\pi\bar{\tau}^{pq}(\lambda_p,\lambda_q)_-
+2\pi i(\xi^p+\phi^p,\,\lambda_p)
\Bigr]
\label{SNtheta}
\end{multline}
where $(a,b):=K_{\alpha\beta}a^{\alpha}b^{\beta}$,
$(a,b)_{\pm}:=(P_{\pm}a,\,P_{\pm}b)$ and $\gamma\in (\Lambda^*/\Lambda)^{\otimes g}$.

We want to express the Gaussian sum
\begin{equation}
S(\ell):=\sum_{\omega \in \mathcal{C}_{\Lambda}}\chi^*_{c^1,c_2}(\omega) e^{-\frac{\pi}{2} g_K(\omega,\omega)
+2\pi i \omega_K(\omega,\ell)}
\label{sum1}
\end{equation}
in terms of the Siegel-Narain theta function \eqref{SNtheta}.
Here $\chi_{c^1,c_2}(\omega)$ is a quadratic refinement \eqref{B:chi} of the form $\omega_K$, i.e.
\begin{equation*}
\chi(\omega_1+\omega_2)=\chi(\omega_1)\chi(\omega_2)
e^{i\pi \omega_K(\omega_1,\omega_2)}.
\end{equation*}
In the basis introduced in section~\ref{sec:basis} the metric $g_K(\omega,\omega)$
has the form ($\omega=m^1_p\alpha^p+m^p_2 \beta_p$)
\begin{equation*}
g_K(\omega,\omega)=(m_2-m^1\cdot\bar{\tau})\cdot(\tau_2^{-1}\otimes \mu)\cdot
(m_2-\tau \cdot m^1).
\end{equation*}
Thus the sum \eqref{sum1} takes the form
\begin{equation}
S(\ell)=\sum_{m_2,m^{1}\in\Zh^{Ng}}
e^{-\pi(m_2-m^1\cdot\bar{\tau})\cdot(\tau_2^{-1}\otimes \mu)\cdot
(m_2-\tau \cdot m^1)
-i\pi (m^1_p,m_2^p)
+2\pi i(m^1_p,\bar{\ell}_2^p)-2\pi i(m_2^p,\bar{\ell}^1_p)
}
\label{S1}
\end{equation}
where $\bar{\ell}^{1}=\ell^{1}+c^1$
and $\bar{\ell}_{2}=\ell_{2}+c_2$.
Using
\begin{equation}
\sum_{m^{I}\in\Zh}e^{-\pi m^{I}m^{J}A_{IJ}
+2\pi B_{I}m^{I}}
=(\det A_{IJ})^{-1/2}\sum_{t_{I}\in\Zh}
e^{-\pi(t_{I}+iB_{I})A^{IJ}(t_{J}+iB_{J})}.
\label{PSF}
\end{equation}
we do Poisson resummation with respect to $m_2$ in \eqref{S1}:
\begin{multline*}
S=\frac{(\det2\tau_2)^{N/2}}{(\det\mu)^{g/2}}\,
e^{-2\pi\,\bar{\ell}^1\cdot(\tau_2\otimes\mu)\cdot\bar{\ell}^1}
\,\sum_{m^1,t\in\Zh}
e^{2\pi i m^1\cdot T\cdot \bar{\ell}^1-4\pi  t\cdot (\tau_2\otimes \Gamma)\cdot \bar{\ell}^1
+2\pi i m^1\cdot(\mathbbmss{1}_g\otimes K)\cdot \bar{\ell}_2}
\\
\times
e^{
i\pi m^1\cdot T\cdot m^1
+2\pi i\, t\cdot (\mathbbmss{1}_g\otimes K^{-1})\cdot T\cdot m^1
-2\pi t\cdot (\tau_2\otimes \mu^{-1})\cdot t
}.
\end{multline*}
where $T=-\tau_1\otimes K+i\tau_2\otimes\mu=-\bar{\tau}\otimes KP_+
-\tau\otimes KP_-$.
Representing $\mu^{-1}=\Gamma K^{-1}$,
$1=P_++P_-$ and $\Gamma=P_+-P_-$ we can rewrite this sum as
\begin{subequations}
\begin{equation}
S(\ell)=\frac{(\det2\tau_2)^{N/2}}{(\det \mu)^{g/2}}\,
e^{-2\pi\,\bar{\ell}^1\cdot(\tau_2\otimes\mu)\cdot\bar{\ell}^1}
\!\!\sum_{m^1,t\in\Zh}e^{
-i\pi\, p_{L}\cdot(\bar{\tau}\otimes\mu)\cdot p_{L}
+i\pi\, p_{R}\cdot(\tau\otimes \mu)\cdot p_{R}
+2\pi i p_L\cdot(\mathbbmss{1}\otimes \mu)\bar{\ell}_{\bar{\tau}}
-2\pi i p_R\cdot(\mathbbmss{1}\otimes \mu)\bar{\ell}_{\tau}
}
\label{S2}
\end{equation}
where
\begin{alignat}{2}
p_{L}&=\bigl[m^1+\mathbbmss{1}_g\otimes K^{-1}t\bigr]_+
+\bigl[\mathbbmss{1}_g\otimes K^{-1}t\bigr]_-,
&\quad
p_{R}&=\bigl[m^1+\mathbbmss{1}_g\otimes K^{-1}t\bigr]_-
+\bigl[\mathbbmss{1}_g\otimes K^{-1}t\bigr]_+;
\\
\bar{\ell}_{\tau}&=\bar{\ell}_2-\tau \bar{\ell}^1,
&\qquad
\bar{\ell}_{\bar{\tau}}&=\bar{\ell}_2-\bar{\tau} \bar{\ell}^1.
\label{psipsi}
\end{alignat}
\end{subequations}
Now $t$ can be uniquely
written as $t=(\mathbbmss{1}_g\otimes K)\cdot s+\delta$.
We can associate the element $\gamma=(\mathbbmss{1}_g\otimes K^{-1})\cdot\delta$
of $(\Lambda^*/\Lambda)^{\otimes g}$
to $\delta$.
So one sees that $\{p_{L}^+,p_{R}^-\}$ and $\{p_{L}^-,p_{R}^+\}$
form two independent lattices. Using that
$\mu P_+=K P_+$ and $\mu P_-=-K P_-$ one sees that
these sums exactly correspond to the sum in the Siegel-Narain
theta function \eqref{SNtheta}.
Using relations $\mu=K\Gamma$ and $\Gamma^2=1$ one obtains that $\det\mu=|\det K|$.
Finally the sum \eqref{S1} splits as follows:
\begin{equation}
S(\ell)=\frac{(\det 2\tau_2)^{N/2}}{|\det K|^{g/2}}\,
e^{2\pi i\omega_K(\ell^{1,0},\ell^{0,1})}\,
\sum_{\gamma\in(\Lambda^*/\Lambda)^{\otimes g}}
\Theta^{c^1,-c_2}_{\Lambda+\gamma}\bigl(\tau,P_{\pm};\,-\ell^{1,0}\bigr)
\Theta^{c^1,c_2}_{\Lambda+\gamma}\bigl(-\bar{\tau},P_{\pm};\,\ell^{0,1}\bigr)
\label{B:result}
\end{equation}
where $\ell^{0,1}=(\ell_2-\bar{\tau}\ell^1)_++(\ell_2-\tau \ell^1)_-$
and  $\ell^{1,0}=(\ell_2-\tau\ell^1)_++(\ell_2-\bar{\tau} \ell^1)_-$.

\newpage


{\small
\begin{thebibliography}{99}
\bibitem{Witten:2003ya}
E.~Witten,
``SL(2,Z) action on three-dimensional conformal field theories with  Abelian
symmetry,''
arXiv:hep-th/0307041.

\bibitem{Gukov:2004id}
S.~Gukov, E.~Martinec, G.~Moore and A.~Strominger,
``Chern-Simons gauge theory and the AdS(3)/CFT(2) correspondence,''
arXiv:hep-th/0403225.

\bibitem{DW}
  R.~Dijkgraaf and E.~Witten,
  ``Topological Gauge Theories And Group Cohomology,''
  Commun.\ Math.\ Phys.\  {\bf 129}, 393 (1990).

\bibitem{Freed:1992vw}
  D.~S.~Freed,
  ``Classical Chern-Simons theory. Part 1,''
  Adv.\ Math.\  {\bf 113}, 237 (1995)
  [arXiv:hep-th/9206021];
  ``Classical Chern-Simons theory. Part 2,''
  http://www.ma.utexas.edu/users/dafr/cs2.pdf


\bibitem{Conway} J.H.~Conway and N.J.A.~Sloane, \textit{Sphere packing, lattices
and groups}, third edition, Springer

\bibitem{Nikulin} V.V.~Nikulin, \textit{Integral symmetric
bilinear forms and some of their applications},
Math. USSR Izvestija Vol. 14 (1980), No. 1, p. 103

\bibitem{Manoliu}
M.~Manoliu, ``Abelian Chern-Simons theory'',
J.Math.Phys. 39 (1998) 170-206; dg-ga/9610001.



\bibitem{Read}N. Read, ``Excitation structure of the hierarchy scheme  in the
fractional quantum Hall effect,'' Phys. Rev. Lett. {\bf 65}(1990)1502

\bibitem{Frohlich:1991wb}
J.~Frohlich and A.~Zee,
``Large scale physics of the quantum Hall fluid,''
Nucl.\ Phys.\ B {\bf 364}, 517 (1991).

\bibitem{Wen} B.~Block and X.G.~Wen,
Phys. Rev. B \textbf{42} 8133--8144; \textbf{42} 8145--8156;
\textbf{43}, 8337--8349.

\bibitem{Zee}
A.~Zee, ``Quantum Hall Fluids'', cond-mat/9501022

\bibitem{Frohlich:1994nk}
  J.~Frohlich, U.~M.~Studer and E.~Thiran,
  ``Quantum theory of large systems of nonrelativistic matter,''
  Lectures given at NATO Advanced Study Institute: Les Houches Summer School, Session 62, Aug 1994,
arXiv:cond-mat/9508062.

\bibitem{Stipsicz}
A.I.~Stipsicz, ``On the vanishing of the third spin cobordism group $\Omega\sb 3\sp {\rm Spin}$'',
Zap. Nauchn. Sem. S.-Peterburg. Otdel. Mat. Inst. Steklov. (POMI) 267 (2000),
Geom. i Topol. 5, 290--302, 332; translation in J. Math. Sci. (N. Y.) 113 (2003), no. 6, 898--905.

\bibitem{Jenquin} J.~Jenquin, ``Spin TQFTs and Chern-Simons Gauge Theory'' (PhD thesis),
math.DG/0504524

\bibitem{Dai:1994kq}
  X.~Z.~Dai and D.~S.~Freed,
  ``eta invariants and determinant lines,''
  J.\ Math.\ Phys.\  {\bf 35}, 5155 (1994)
  [Erratum-ibid.\  {\bf 42}, 2343 (2001)]
  [arXiv:hep-th/9405012].

\bibitem{Witten:1996hc}
E.~Witten,
``Five-brane effective action in M-theory,''
J.\ Geom.\ Phys.\  {\bf 22}, 103 (1997)
[arXiv:hep-th/9610234].

\bibitem{Diaconescu:2003bm}
E.~Diaconescu,  D.~S.~Freed, and G.~Moore
``The M-theory 3-form and E(8) gauge theory,''
arXiv:hep-th/0312069.

\bibitem{mumford} D.~Mumford, \textit{Tata Lectures on Theta I},
Birkh\"auser, 1983.

\bibitem{Atiyah} M.~Atiyah, ``Riemann surfaces and spin structures'',
\textit{Ann. Sci. \'{E}cole Norm. Sup.}~\textbf{4} (1971), 47 -- 62.

\bibitem{CAV} C.~Birkenhake and H.~Lange, \textit{Complex Abelian Varieties},
second edition, A series of Comprehansive Studies in Mathematics, Springer 2004

\bibitem{Alvarez-Gaume:1987vm}
  L.~Alvarez-Gaume, J.~B.~Bost, G.~W.~Moore, P.~Nelson and C.~Vafa,
  ``Bosonization On Higher Genus Riemann Surfaces,''
  Commun.\ Math.\ Phys.\  {\bf 112}, 503 (1987).


\bibitem{Milnor} J.~Milnor and D.~Husemoller, \textit{Symmetric bilinear forms},
Springer-Verlag, New York,
1973, Ergebnisse der Mathematik und ihrer Grenzgebiete, Band 73. MR 58 \#22129

\bibitem{HopkinsSinger}  M.J.~Hopkins, I.M.~Singer,
``Quadratic functions in geometry, topology,and M-theory'',
math.AT/0211216

\bibitem{Milgram} R.J.~Milgram, ``Surgery with coefficients'',
Ann.~of~Math.~\textbf{100} (1974) 194--248.

\bibitem{Gegenberg:1993gd}
J.~Gegenberg and G.~Kunstatter,
``The Partition function for topological field theories,''
Annals Phys.\  {\bf 231}, 270 (1994)
[arXiv:hep-th/9304016].



\end{thebibliography}
}
\end{document}